\documentclass[a4paper,11pt]{article}
\usepackage{pos}
\usepackage{mathbbol}
\usepackage{amsfonts}
\usepackage{longtable}
\usepackage{hhline}
\usepackage{bm}

\title{Measurement of Quadrupole Deformation using E$2$ and M$1+$E$2$ Transitions in Heavy Isotopes in the Mass Range of $150<A<250$}
\ShortTitle{Quadrupole Deformation using E$2$ and M$1+$E$2$ Transitions}

\author*[a]{Prajwal MohanMurthy}\emailAdd{prajwal@alum.mit.edu}
\author[b]{Lixin Qin$^{\dagger,}$\footnote[0]{$^{\dagger}$Co-first author}}
\author[c]{Jeff A. Winger}\emailAdd{j.a.winger@msstate.edu}

\affiliation[a]{Laboratory for Nuclear Science, Massachusetts Institute of Technology\\
77 Mass. Ave., Cambridge, MA 02139, USA}

\affiliation[b]{Wellesley College,\\
106 Central St., Wellesley, MA 02481, USA}

\affiliation[c]{Department of Physics and Astronomy, Mississippi State University\\
PO Box 5167, Mississippi State, MS 39762, USA}

\abstract{\scriptsize
The measurement of a permanent electric dipole moment (EDM) in atoms is crucial for understanding the origins of CP-violation. Quadrupole and octupole deformed nuclei exhibit a significantly enhanced atomic EDM. However, accurate interpretation of the EDM in such systems requires the characterization of their deformation. While nuclear deformation is indicated in various structure models, there is substantial mutual disagreement between the theoretical models or between theoretical models and experimental values. Experimental confirmation of the same, particularly in heavy isotopes essential for EDM measurements, is lacking.
\\
Nuclear E$2$ transitions allow access to quantify quadrupole deformation, but these transitions are often mixed with M$1$ transitions. Both E$2$ and M$1$ transitions are well characterized by Weisskopf estimates, which rely on a single-particle approximation. However, deviations from measurements arise due to collective nuclear deformation. To utilize pure E$2$ and mixed E$2$+M$1$ transition lifetimes for determining quadrupole deformation, accurate Weisskopf estimates for heavy nuclei are essential. Previously, Weisskopf estimates were only available for the mass range $A<150$, and in this work we have extended the Weisskopf estimates of both E$2$ and M$1$ transition lifetimes to the mass range of $150\le A\le 250$.
\\
This allowed us to comprehensively study the deviation of E$2$ and M$1+$E$2$ transition lifetimes from the newly established Weisskopf estimates in $91$ candidate isotopes, by comparing the transition lifetimes of the candidate isotopes to their nearest even-even counterparts, whose E$2$ transition strengths are very well understood. Estimates of collective nuclear quadrupole deformation in $67$ of these isotopes were obtained, either from E$2$ or M$1+$E$2$ transition lifetimes, and in $32$ cases they were obtained from both types of transitions independently. We find that the quadrupole deformation extracted from the two different types of transitions are mutually consistent, as well as that they follow the trends established in theory. We thereby identify the isotopes $^{223,225}$Fr, $^{221,223}$Ra, $^{223,225,227}$Ac and $^{229}$Pa, where EDM measurements are foreseen and information on nuclear deformation is needed, for which no measurement of nuclear quadrupole deformation has been made.}

\FullConference{10$^{th}$ International Conference on Quarks and Nuclear Physics (QNP2024)\\
8-12 July, 2024\\
Barcelona, Spain\\}

\begin{document}
\maketitle

\section{Introduction}

The amount of charge-parity (CP) violation in the Standard Model, arising from the weak sector, is insufficient \cite{Riotto1999-vo} to explain the observed baryon asymmetry of the universe \cite{Morrissey2012-tt,Aghanim2020-cm}. Measurement of a statistically significant CP violating electric dipole moment (EDM) in multiple systems, not only provides a means to quantify the amount of CP violation from various sources \cite{Chupp2015-ns}, but also directly probes the CP violation arising from the strong sector \cite{t_Hooft1976-lq}.

Even though measurement of the EDM for the neutron was the first to be pioneered \cite{Kirch2020-dr,Mohanmurthy2021-ou}, atoms and molecules have emerged as lucrative systems in which to measure CP violating parameters of the atomic EDM \cite{Engel2013-rk,Chupp2019-cm} and the nuclear magnetic quadrupole moment (MQM) \cite{Flambaum1994-fp,Flambaum2014-qn}. The EDM of atoms with quadrupole and octupole deformed nuclei are dramatically enhanced \cite{Sushkov1984-cb,Auerbach1996-sp,Spevak1997-bg,Dzuba2002-or,Khriplovich2012-if,Flambaum2014-qn}. Nuclear quadrupole deformation is a prerequisite for this enhancement of the EDM, both via the nuclear MQM \cite{Lackenby2018-ua,Flambaum2022-kj}, as well as the nuclear Schiff moment \cite{Auerbach1996-sp, Spevak1997-bg}. The particulars of the enhancement have been described in further detail in \emph{Refs.}~\cite{Mohanmurthy2020-np,MohanMurthy2024-dh}. In this work, we are focused on nuclear quadrupole deformation, accessed via nuclear $\gamma$-ray spectroscopy measurements, that are already available.

The nuclear quadrupole moment is characterized by $\beta_2$, which is the coefficient associated with the quadrupole term of the multipole expansion defining the surface of a nucleus given by \cite{Leander1986-fr}
\begin{equation}
R=c_V R_A \left(1+\sum_{\lambda=2}\beta_{\lambda} Y^{m=0}_{\lambda}\right),~\label{eq2024-1-1}
\end{equation}
where $c_V=1-\{(1/\sqrt{4\pi})\sum_{\lambda=2}\beta^2_{\lambda}\}$ is the volume normalization such that $R_A=R_0 A^{1/3}=1.2~\text{fm}\cdot A^{1/3}$, $A$ is the total number of nucleons in the nucleus, $\beta_{\lambda}$ are the $2^{\lambda}$-pole structure deformation coefficients, and $Y^m_{\lambda}$ are spherical harmomic functions. The quadrupole deformation of even-even nuclei has been well vetted \cite{Pritychenko2016-tw,Pritychenko2017-pm}, and can be accessed in the NNDC database \cite{nndc}.

Nuclear quadrupole deformation and shape collectivity have been well measured \cite{Garg2023-jx, Heyde2011-mc} for a large range of nuclei; however, the same are only scantly known for isotopes with $A\!>\!200$ \cite{Dracoulis2016-uf}. Especially since the octupole deformed region around $A\!\sim\!223$ is of great interest to the EDM community \cite{MohanMurthy2024-dh}, it is absolutely necessary to characterize the deformation of these isotopes. Nuclear theory models which predict the deformation are mature \cite{Agbemava2016-rx,Ebata2017-nt,Moller2016-gt}. Nonetheless, there are inconsistencies between the various theoretical models themselves, \emph{e.g.}~for $^{224}$Ra, $\beta_2=0.177$ in \emph{Ref.}~\cite{Agbemava2016-rx}, $\beta_2=0.18$ in \emph{Ref.}~\cite{Ebata2017-nt}, and $\beta_2=0.143$ in \emph{Ref.}~\cite{Moller2016-gt}. There are also inconsistencies between measurements and theoretical models, \emph{e.g.}~the measurements indicate that the quadrupole deformation of $^{224}$Ra is $\beta_2=0.2022(25)$ \cite{Pritychenko2016-tw}. Precisely knowing the deformation is critical to interpreting the EDM in terms of more fundamental sources of CP violation arising from individual nucleons \cite{Flambaum1994-fp,Lackenby2018-ua} and their mutual interactions \cite{Engel2013-rk,Chupp2015-ns}. This provides additional motivation for explicitly measuring the deformation, as opposed to relying solely on the theoretical models.

\section{Electromagnetic Transition Lifetimes}

In order to better understand the nuclear transition energy spectra, we began by studying the Weisskopf estimates \cite{Blatt1979-vo,Blatt2012-fk} for single particle electromagnetic transitions in heavy isotopes, $150<A<250$. The Weisskopf estimates give ballpark estimates of the decay width for nuclear transitions under the assumption that a single nucleon is participating in the process \cite{Weisskopf1957-sa}. The Weisskopf estimates further assume that the initial and final states of the single particle participating in the transition has an angular momentum of $l=1/2$. This assumption is relaxed by incorporating multiplicity into the Weisskopf estimates, as done by the Moszkowski estimates \cite{Moszkowski1953-ok}. %In this work, we carefully consider a more generalized version of the Weisskopf estimates with multiplicity, but simply refer to it as Weisskopf estimates.

%Eq.
Nuclear transitions can be electric or magnetic in type, where the single nucleon participating in the transition interacts with the electric or magnetic fields (arising from the core of the nucleus), respectively. These transitions are characterized by the inverse of the decay lifetime ($\tau$), or the width ($\Gamma$). In order to remove the dependence on the energy of the photon emitted ($\mathcal{E}_{\gamma}$), it is convenient to use strengths of the transition, $B(i\lambda)$, defined by \cite{Krane1987-lw}
\begin{eqnarray}
\frac{\text{Ln}(2)}{\tau_{1/2}(i\lambda)} &=& \Gamma(i\lambda) = \frac{8\pi(\lambda+1)}{\hbar\lambda\left(\left(2\lambda+1\right)!!\right)^2}\left(\frac{\mathcal{E}_{\gamma}}{\hbar c}\right)^{2\lambda+1}\cdot B_{\text{total}(i\lambda)},~\nonumber\\
B_{\text{total}(i\lambda)}&=&B_{w:i\lambda}+B_{\beta}(i\lambda)+ \mathcal{O},~\label{eq2024-1-2}
\end{eqnarray}
where $i\in\{E,M\}$, $B_{w}$ are single particle Weisskopf estimates, $B_{\beta}$ are the strengths due to collective deformation, $\tau_{1/2}$ is the half-life of the transition, which is closely linked to the lifetime, $\tau$, as $\tau_{1/2}=\text{Ln}(2)\tau$ as well as the transition width, $\Gamma(i\lambda)$, as $\Gamma(i\lambda)=1/\tau$, and $\mathcal{O}$ encompasses any N-body contributions \cite{Miyagi2024-iy,Stroberg2022-uk}. In other words, transition lifetimes and their respective transition strengths are interchangeable using Eq.~\ref{eq2024-1-2}. In Eq.~\ref{eq2024-1-2}, the units of the strength of the transition are $\mathbb{e}^2\cdot \text{fm}^{2\lambda}$ or $\mu_N^2\cdot \text{fm}^{2\lambda-2}$, for $2^\lambda$-pole electric or magnetic type transitions, respectively. The single particle Weisskopf estimates of the width for the electric and magnetic transitions then can be written as \cite{Krane1987-lw}
\begin{eqnarray}
B_{w:E\lambda} &=& R_0^{2\lambda}\frac{1}{4\pi}\left(\frac{3}{\lambda+3}\right)^2 A^{2\lambda/3} \mathbb{e}^2,~\label{eq2024-1-3}\\
B_{w:M\lambda} &=& R_0^{2\lambda-2}\frac{10}{4\pi}\left(\frac{3}{\lambda+3}\right)^2 A^{(2\lambda-2)/3} \mu_N^2,~\label{eq2024-1-4}
\end{eqnarray}
where $\mathbb{e}$ is the elementary charge (with units given by $\mathbb{e}^2=1.44~\text{MeV}\cdot\text{fm}$), and $\mu_N$ is the nuclear magneton (with units given by $\mu_N=0.10515446~\mathbb{e}\cdot\text{fm}$ or $\mu_N^2=0.0159~\text{MeV}\cdot\text{fm}^3$).

The nuclear electromagnetic potentials can be written in terms of multi-pole moments, as described in \emph{Ref.}~\cite{MohanMurthy2024-dh}. When the photon emitted in the transition arises due to the interaction with a specific electromagnetic $2^\lambda-$pole moment, the transition is typically labeled as E$\lambda$ or M$\lambda$, where `E' or `M' represent the electric or magnetic transition, and $\lambda$ refers to the $2^\lambda-$pole moment underlying the transition. Emission of a photon can naturally lead to a change in the spin of the system by $\Delta j=1$, but the spin can also change by any integer, $\Delta j=\mathbb{Z}$, in combination with the change of angular momentum of the system. The change in parity is defined by the type of $E\lambda$ or M$\lambda$ (Table 1 of \emph{Ref.}~\cite{MohanMurthy2024-dh}). The allowed change in spin and parity of the system undergoing an electromagnetic transition is summarized in Table~\ref{tab2024-1-1}.
%Table

\begin{table}[t]
\centering
\caption[]{Selection rules for spin ($\Delta j$) and parity ($\Delta \pi=\pm$ indicates even and odd, respectively) due to E$\lambda$ (Electric) and M$\lambda$ (Magnetic) transitions, along with their label described by $2^\lambda$-pole.}
\label{tab2024-1-1}
\begin{tabular}{c r c c}
\hline
$\bm{\lambda}$ & {\bf Label} & {\bf E} & {\bf M} \\
\hline
\hline
1 & Di-pole & $\{|\Delta j| = 1,~\Delta \pi = -\}$ & $\{|\Delta j| = 1,~\Delta \pi = +\}$ \\
2 & Quadru-pole & $\{|\Delta j| = 2,~\Delta \pi = +\}$ & $\{|\Delta j| = 2,~\Delta \pi = -\}$ \\
3 & Octu-pole & $\{|\Delta j| = 3,~\Delta \pi = -\}$ & $\{|\Delta j| = 3,~\Delta \pi = +\}$ \\
\hline
\hline
\end{tabular}
\end{table}

%\subsection{Transition Lifetimes due to Collective Deformation}
%due to nuclear deformation

Quadruple deformation is ubiquitous across the nuclide chart \cite{MohanMurthy2024-dh}. The Weisskopf estimates only give a ballpark measure of the transition lifetime, mostly because it is dominated by contributions from nuclear deformation, indicated by $B_{\beta}$ in Eq.~\ref{eq2024-1-2}. As one might expect, the transition strength of the electric E$\lambda$ transition is directly linked to the $2^\lambda$-pole nuclear deformation \cite{Wollersheim1993-ro,Butler1996-px}
\begin{eqnarray}
B_{\beta}(E\lambda: j_i \rightarrow j_f) &=& \frac{2 \lambda+1}{16\pi}\left\Vert\langle j_i,0;\lambda,0|j_f,0\rangle\right\Vert^2 Q^2_{\lambda,0},~\label{eq2024-1-5}
\end{eqnarray}
where the inner product is the Clebsch-Gordan coefficient, and $Q_{\lambda,0}$ is the electric $2^\lambda$-pole moment of the nucleus given by \cite{Leander1988-ww}
\begin{eqnarray}
Q_{\lambda\ge2,0}=\frac{3}{\sqrt{(2\lambda+1)\pi}}Z\mathbb{e}R^{\lambda}_A \bar{\beta}_\lambda.~\label{eq2024-1-6}
\end{eqnarray}
To the first order $\bar{\beta}_\lambda \approx \beta_\lambda$, where $\beta_\lambda$ are defined in Eq.~\ref{eq2024-1-1}, and $Z$ is the number of protons in the nucleus. However, higher order corrections to $\bar{\beta}_\lambda$ are available in \emph{Ref.}~\cite{Leander1988-ww}.

In this way measuring the E$2$ transition lifetime of the $\gamma$-ray decay in Eq.~\ref{eq2024-1-2}, and correcting for the single particle-contribution to the lifetime dictated by the Weisskopf estimate in Eq.~\ref{eq2024-1-3}, one can gain access to the underlying nuclear deformation directly via Eqs.~\ref{eq2024-1-5} and \ref{eq2024-1-6}. However, while pure E$2$ transitions are readily observed, it is relative rare to observe an E$3$ transition within a nucleus, owing to diminishing rate with higher order transitions \cite{nndc}. Also, pure E$2$ transitions are sometimes mixed with M$1$ transitions. Since nuclear deformation does not affect M$1$ transitions, M$1$+E$2$ transition lifetimes can also be used to characterize nuclear deformation after correcting for the Weisskopf estimate for the M$1$ transition, in addition to that for the E$2$ transition.

\section{Extending Weisskopf Estimates in the Range $150<A<250$}

The Weisskopf estimates for M$1$ and E$2$ transitions have been extensively studied for lighter isotopes. A global analysis of these estimates maybe found in \emph{Refs.}~\cite{Endt1979-sz, Endt1993-fc, Endt1981-wp} for $A\in[4,150]$. In this section, we will attempt to extend these single particle estimates over a range $150<A<250$, which will help us experimentally confirm the quadrupole deformation in isotopes relevant for EDM measurements, from existing E$2$ and M$1$+E$2$ transition lifetimes \cite{nndc}.

\subsection{Weisskopf Estimates for Pure M1 and E2 Transitions}

For obtaining Weisskopf estimates, we choose to use nuclei with spin-$0$ ground state, given that such systems provide for the simplest case. In the case of spin-$0$ nuclei, higher order electromagnetic multi-polar corrections are absent \cite{Siegbahn1965-le,Stech1952-np}. We further exclude nuclei with non-band transitions or inter-band transitions, and consider candidate isotopes with pure M$1$ and E$2$ transitions directly to the ground state, so that the transition energy is adequately represented by the energy of the emitted photon in Eq.~\ref{eq2024-1-2}. In addition, since E$2$ transition strength is strongly dependent on nuclear quadrupole deformation, we also selected the isotopes which are known to be minimally quadrupole deformed ($\beta_2<0.1$) in finite range droplet macroscopic model (FRDM) calculations \cite{Moller2016-gt}, such that the contribution from quadrupole deformation to the E$2$ strength is small (Eqs.~\ref{eq2024-1-2},\ref{eq2024-1-5}). However, such a filter involving the theoretical quadrupole deformation was not applied to the list of isotopes with M$1$ transitions since M$1$ transition strength is not directly affected by nuclear quadrupole deformation. These pre-conditions severely constrain the available measurements in the mass range of $150<A<250$.

\begin{figure}[p]
\centering
%\begin{tabular}{cc}
    %\multirow{-13.5}{*}{\includegraphics[width=0.44\textwidth]{beta23.png}} & \includegraphics[width=0.54\textwidth]{beta2-MN.png} \\
%  &  \includegraphics[width=0.54\textwidth]{beta3-MN.png}
%\end{tabular}
\includegraphics[width=\textwidth]{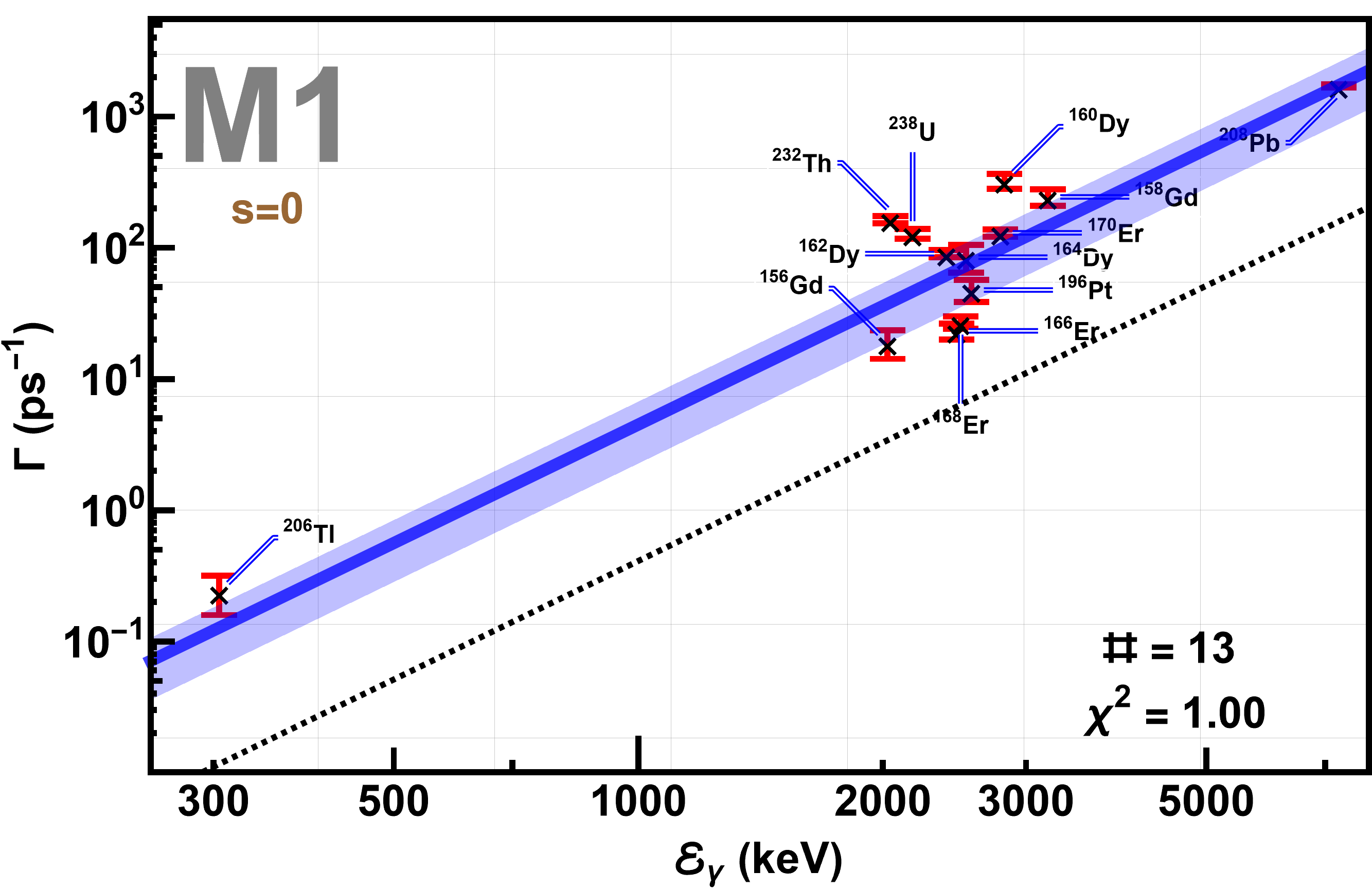}
\includegraphics[width=\textwidth]{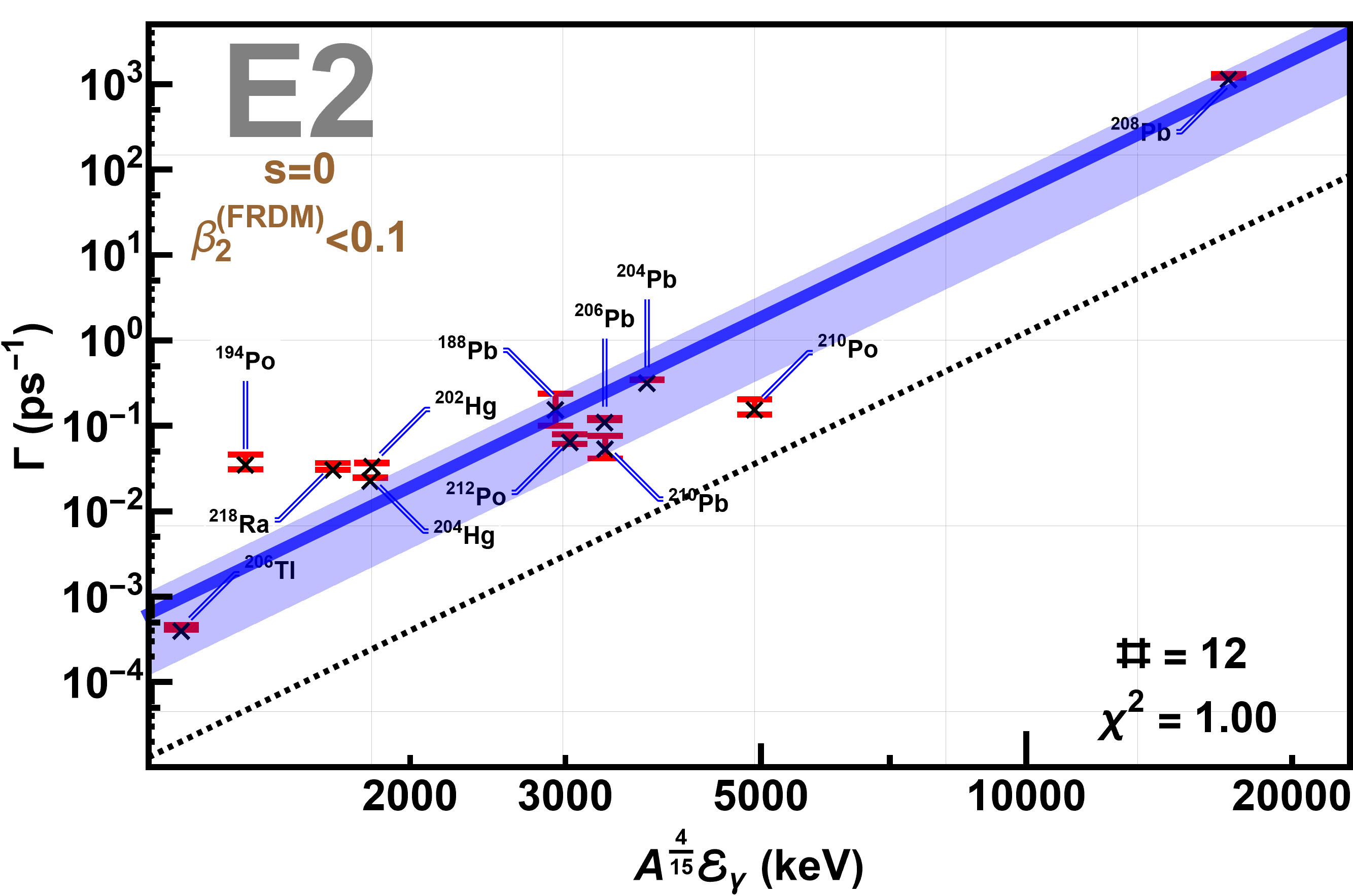}
\caption[]{Plot showing the pure M$1$ (E$2$) transition decay widths as a function of transition energy, $\mathcal{E}_{\gamma}$ ($A^{4/15}\mathcal{E}_{\gamma}$), in the Top (Bottom) panel for $13$ ($12$) candidate isotopes with spin a $0$ ground state. In the case of E$2$ transition decay widths, in the Bottom panel, the candidate isotopes were minimally quadrupole deformed, $\beta^{\text{(FRDM)}}_2<0.1$, according to \emph{Ref.}~\cite{Moller2016-gt}.}
\label{fig2024-1-1}
\end{figure}

In order to obtain the Weisskopf estimates, we used $13$ candidate nuclei for M$1$ transitions and $12$ candidate nuclei for E$2$ transitions, all of which satisfy the above requirements. The energy difference between the nuclear levels involved in the $\gamma$-ray transition ($\mathcal{E}_{\gamma}$) has been plotted as a function of their transition widths in Figure~\ref{fig2024-1-1}, for candidate isotopes with both M$1$ and E$2$ transitions. According to Eqs.~\ref{eq2024-1-2}-\ref{eq2024-1-4}, we fitted the transition width to, $\mathcal{E}^3_{\gamma}$ for M$1$ transitions, and $\mathcal{E}^5_{\gamma}$ for E$2$ transitions. We thus obtained weighted fits for M$1$ and E$2$ transitions, respectively, of
\begin{eqnarray}
\Gamma_{w:\text{M}1}(\mathcal{E}_{\gamma})&=&\underbrace{2.3(9)\times 10^{-9}}_{m_{w:\text{M}1}}\left(\frac{\mathcal{E}_{\gamma}}{\text{keV}}\right)^3~\text{ps}^{-1},~\label{eq2024-1-7}\\
\Gamma_{w:\text{E}2}(\mathcal{E}_{\gamma})&\lesssim&\underbrace{6.0(4)\times 10^{-19}}_{m_{w:\text{E}2}}A^{\frac{4}{3}}\left(\frac{\mathcal{E}_{\gamma}}{\text{keV}}\right)^5~\text{ps}^{-1},\label{eq2024-1-8}
\end{eqnarray}
where the uncertainty of the slope represents the $68.3$\% C.I. of the standard error on the mean, such that the reduced-$\chi^2\approx1$, and $m_i$ represents the respective slopes. Typically the relative uncertainty of the measured transition lifetime dominated, compared to that of the photon energy. Since the candidate list is limited in Figure~\ref{fig2024-1-1}, we also tested the resilience of the slope values above by checking if removing a single data point changes the values significantly. We note that even though $^{206}$Tl and $^{208}$Pb fall on the extreme sides of both the plots in Figure~\ref{fig2024-1-1}, the removal of these two data points does not impact the slope values in Eqs.~\ref{eq2024-1-7} and \ref{eq2024-1-8} significantly. It is also important to note that the horizontal axes in Figure~\ref{fig2024-1-1}, corresponding to the fit to E$2$ transition widths, involves $A^{4/15}\mathcal{E}_{\gamma}$, so that $\Gamma(\text{E}2)\propto[(A^{4/15}\mathcal{E}_{\gamma})^5=A^{4/3}\mathcal{E}_{\gamma}^5]$ in Eq.~\ref{eq2024-1-8}, yielding a straight line fit in the log-log plot, analogous to the straight line fit to M$1$ transition widths in Figure~\ref{fig2024-1-1}, where $\Gamma(\text{M}1)\propto\mathcal{E}_{\gamma}^3$ in Eq.~\ref{eq2024-1-7}.

\subsection{Weisskopf Estimates for E2 Transitions in Deformed Nuclei}

Quadrupole deformation affects E$2$ transition lifetimes significantly, sometimes by many orders of magnitude. Even though the Weisskopf estimate for E$2$ transitions, obtained in Eq.~\ref{eq2024-1-8}, used only theoretically mildly quadrupole deformed isotopes, it is still an overestimate due to the pervasive quadrupole deformation present in the heavy isotopes in the range of $150<A<250$. The estimate for the E$2$ transition lifetime in Eq.~\ref{eq2024-1-7} has a non-zero contribution from the nuclear quadrupole deformation entering Eq.~\ref{eq2024-1-2} via Eqs.~\ref{eq2024-1-5} and \ref{eq2024-1-6}. We studied the empirically measured E$2$ transition lifetimes as a function of theoretical quadrupole deformation from the FRDM model \cite{Moller2016-gt}, in order to isolate the effects of quadrupole deformation and thereby extract a precise Weisskopf single particle estimate for E$2$ transitions.

\begin{figure}[t]
\centering
%\begin{tabular}{cc}
    %\multirow{-13.5}{*}{\includegraphics[width=0.44\textwidth]{beta23.png}} & \includegraphics[width=0.54\textwidth]{beta2-MN.png} \\
%  &  \includegraphics[width=0.54\textwidth]{beta3-MN.png}
%\end{tabular}
\includegraphics[width=\textwidth]{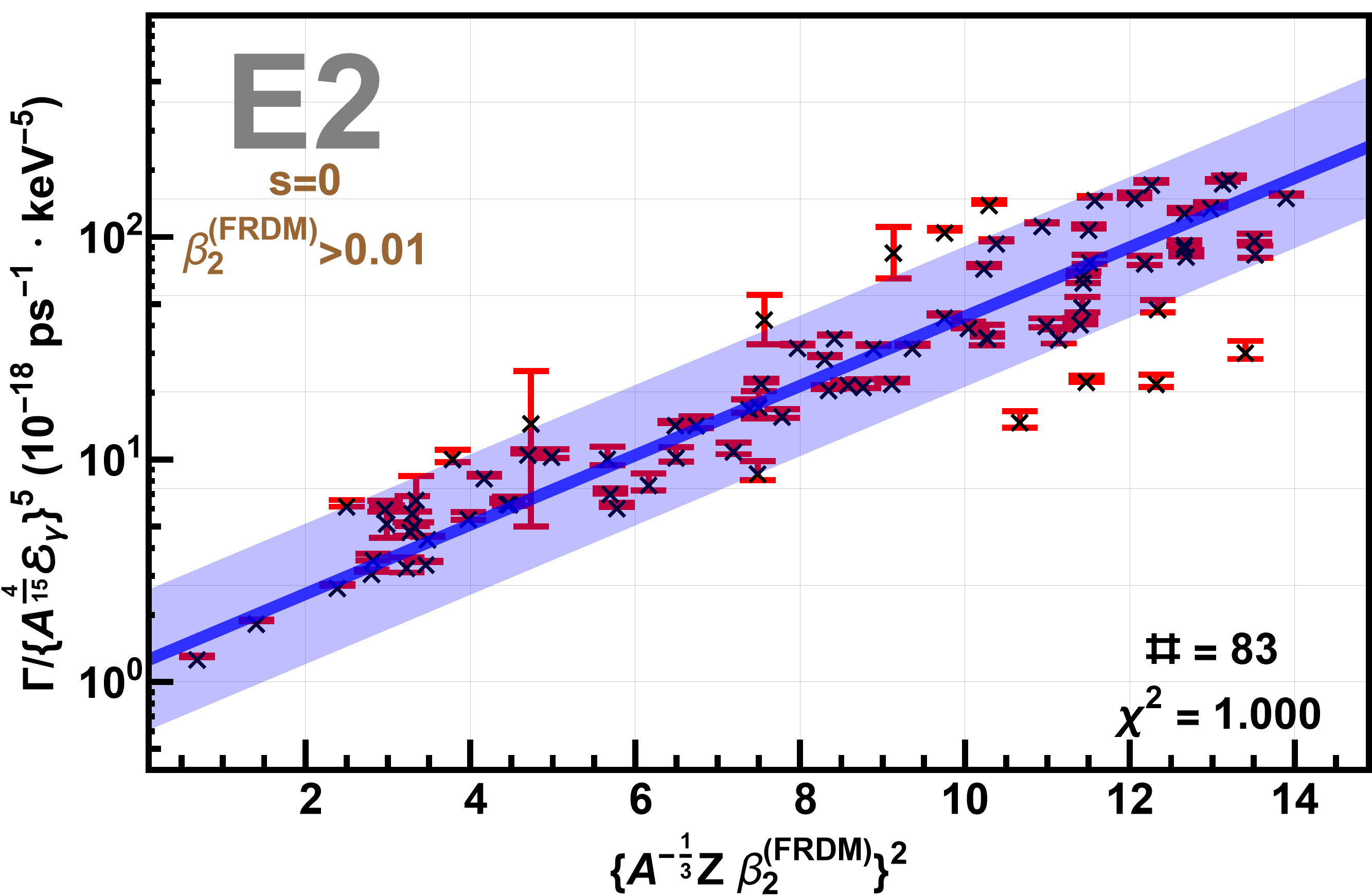}
\caption[]{Plot showing the dependence of the ratio of E$2$ transition widths to transition energy to the power five, $\Gamma/(A^{4/15}\mathcal{E}_{\gamma})^5$, as a function of the square of the theoretical quadrupole deformation coefficient \cite{Moller2016-gt}, $\{\beta^{\text{(FRDM)}}_2\}^2$, as shown in Eq.~\ref{eq2024-1-9}, in order to extract the Weisskopf estimate for the E$2$ transition width for nuclei without any quadrupole deformation, $\beta_2=0$.}
\label{fig2024-1-2}
\end{figure}

By substituting in the nuclear electric quadrupole moment in Eq.~\ref{eq2024-1-6} into the strength of the E$2$ transition in Eq.~\ref{eq2024-1-5}, one can see that the E$2$ transition lifetime (Eq.~\ref{eq2024-1-2}) varies as $\bar{\beta}^2_2$. While $\bar{\beta}_2\approx\beta_2$, to the first order, it is important to note that $\bar{\beta}_2$ has contributions from $\beta_{\lambda>2}$ as well \cite{Leander1988-ww}. The parameter of $\Gamma(\text{E}2)/(A^{4/3}\mathcal{E}_{\gamma}^5)$ was fitted as a function of theoretical values of $\beta^2_2$ in Figure~\ref{fig2024-1-2} which yielded
\begin{eqnarray}
\frac{\Gamma_{w: \text{E}2}/\text{ps}^{-1}}{A^{4/3}(\mathcal{E}_{\gamma}/\text{keV})^5}&=&\underbrace{1.13(5)\times10^{-19}}_{m_{w:\text{E}2}}~+~m_{\beta}\cdot\left(A^{-\frac{1}{3}}Z\beta^{\text{(FRDM)}}_2\right)^2,~\label{eq2024-1-9}
\end{eqnarray}
where the vertical intercept is the slope, $m_{w:\text{E}2}$, associated with the Weisskopf estimate for E$2$ transitions in Eq.~\ref{eq2024-1-8}. Since the contribution of the
quadrupole deformation term to the width of the E$2$ transition in Eq.~\ref{eq2024-1-2}, $B_{\beta}$, depends on the nuclear electric quadrupole moment, $Q_{2,0}^2$, in Eq.~\ref{eq2024-1-5}, it can be seen from Eq.~\ref{eq2024-1-6} that the independent quantity that needed to be employed in Figure~\ref{fig2024-1-2} as the horizontal coordinate is $\{A^{-1/3}Z\beta^{\text{(FRDM)}}\}^2$. Each of the individual parameters of  $\{ A, Z, \beta^{\text{(FRDM)}}_2 \}$ are unique to each candidate isotope. The slope in Eq.~\ref{eq2024-1-9}, $m_{\beta}$, maybe deduced by the combination of Eqs.~\ref{eq2024-1-2},~\ref{eq2024-1-5}, and \ref{eq2024-1-6}.

Clearly, even using mildly quadrupole deformed nuclei in Figure~\ref{fig2024-1-1} yielded an overestimate for the widths in Eq.~\ref{eq2024-1-8}. By removing the pre-condition of minimal theoretical quadrupole deformation, $\beta_2<0.1$, for the candidate nuclei whose E$2$ transition lifetimes were used to obtain the estimate in Eq.~\ref{eq2024-1-8}, the list of candidates increased in Figure~\ref{fig2024-1-2}, which resulted in nearly an order of magnitude more precise estimate for the widths in Eq.~\ref{eq2024-1-9}, compared to in Eq.~\ref{eq2024-1-8}. Furthermore, we used a constraint of $\beta_2>0.01$, simply to eliminate typically very light nuclei where the FRDM model is unable to characterize tiny deformations accurately.

\section{Measurement of Quadrupole Deformation}

By understanding the single particle contributions to the E$2$ and M$1$ transition lifetimes, we were able isolate the contribution from quadrupole deformation to E$2$ and M$1$+E$2$, respectively, as dictated by Eq.~\ref{eq2024-1-2}. In this section, we have considered pure E$2$ as well as M$1$+E$2$ transitions to the ground state. We were able to isolate the contribution of the quadrupole deformation to the transition lifetimes, which allowed us to characterize their quadrupole deformation. %Furthermore we avoided considering transitions between excited spin-states that are yet to be absolutely established (usually indicated in parenthesis in NNDC \cite{nndc}).

%index#, Z, N, A, E2 Transition Energy (keV), Energy's uncertainty, E2 lifetime (ps), lifetime's uncertainty,  PlusMinus[]*, B(E2) with error (inside PlusMinus[], in (eb)^2), ground state spin, theory-beta_2, calculated Exp. beta_2 (inside PlusMinus[]), deviation of Exp from Theory in sigmas

\subsection{Quadrupole Deformation from E2 Transitions}

%spin-orbit coupling, relativistic corrections, ratio method

In the case of pure E$2$ transitions, the contribution from the Weisskopf single particle estimates, linked to the emitted photon energy according to Eq.~\ref{eq2024-1-9}, was first calculated for each candidate isotope. Then the calculated Weisskopf single particle estimate for the transition lifetime was removed from the empirically measured transition lifetime, which are reported in the NNDC database \cite{nndc}, according to Eq.~\ref{eq2024-1-2}. Subtracting the single particle contribution still leaves contributions due to N-body interaction \cite{Miyagi2024-iy,Stroberg2022-uk}, indicated by $\mathcal{O}$ in Eq.~\ref{eq2024-1-2}.

The empirically measured E$2$ transition lifetime corrected for single particle estimates, can be converted directly to the strength of the transition due to quadrupole deformation, via Eq.~\ref{eq2024-1-2}, which gives us access to the quadrupole deformation parameter of $\beta_2$, via Eqs.~\ref{eq2024-1-5} and \ref{eq2024-1-6}. However, this typically leads to overestimation of the strength, which in turn leads to an overestimation of the quadrupole deformation. The contributions from N-body interactions can in fact be quite large. In order to minimize such spurious contributions, we use the ratio of the E$2$ transition lifetime of the candidate isotope with that of the nearest even-even nuclei,
\begin{eqnarray}
&\frac{\tau^{\text{(even-even)}}_{1/2}}{\tau^*_{1/2}} = \frac{\Gamma^*(\text{E}2: j_i \rightarrow j_f)}{\Gamma^{\text{(even-even)}}(\text{E}2: 0^+ \rightarrow 2^+)} \approx \frac{B^*_{w:\text{E}2}(\mathcal{E}^*_{\gamma})+B^*_{\beta}(\text{E}2: j_i \rightarrow j_f)}{B^{\text{(even-even)}}_{w:\text{E}2}(\mathcal{E}^{\text{(even-even)}}_{\gamma})+B^{\text{(even-even)}}_{\text{adopted}-\beta}(\text{E}2: 0^+ \rightarrow 2^+)},~\label{eq2024-1-10}
\end{eqnarray}
where the parameters marked with an asterisk are the candidate isotopes we have considered, and those marked (even-even) are from the nearest even-even nuclei. When using the ratio of E$2$ transition lifetimes in Eq.~\ref{eq2024-1-10}, we also correct the E$2$ transition strength of the nearest even-even nuclei with its corresponding contribution from the Weisskopf estimate using the appropriate energy of the photon linked to the $2^+ \rightarrow 0^+$ transition ($\mathcal{E}^{\text{(even-even)}}_{\gamma}$). The strength associated with the single particle estimates for the E$2$ transition lifetime obtained in Eq.~\ref{eq2024-1-9} was calculated using Eq.~\ref{eq2024-1-2}. The E$2$ transition strength associated with quadrupole deformation of the even-even nuclei was obtained using the corresponding adopted $B$(E$2$) (or adopted values of $\beta_2$), from a combination of measurements and theory, as reported in \emph{Refs.}~\cite{Pritychenko2016-tw,Pritychenko2017-pm}.

\begin{figure}[p]
\centering
\includegraphics[width=0.900\textwidth]{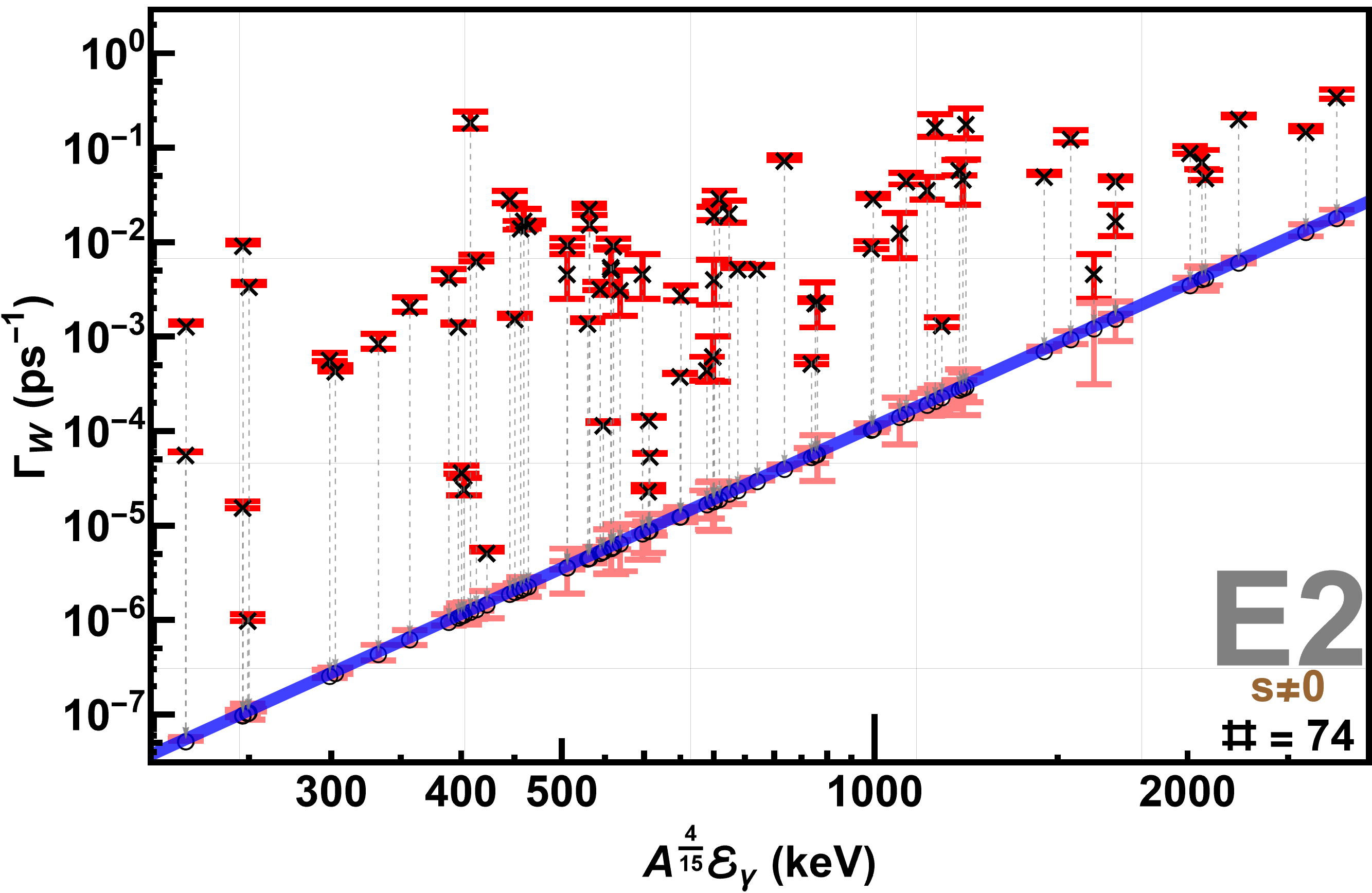}
\includegraphics[width=0.900\textwidth]{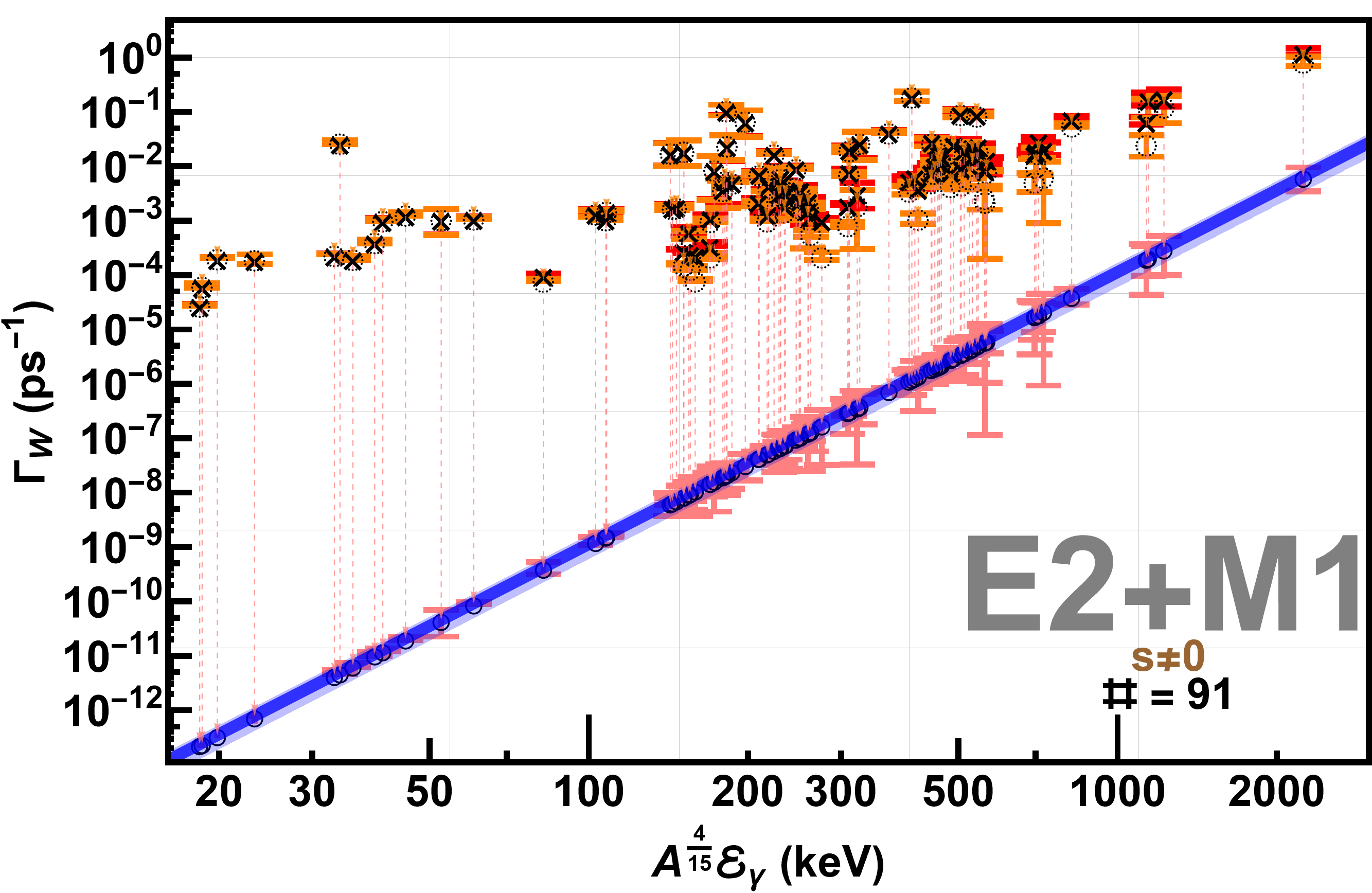}
\caption[]{Top (Bottom): Plot showing the measured E$2$ (M$1+$E$2$) transition widths as a function of transition energy $A^{4/15}\mathcal{E}_{\gamma}$ \cite{Pritychenko2016-tw,Pritychenko2017-pm,nndc}, as in Eq.~\ref{eq2024-1-2}. The central value of the E$2$ (M$1+$E$2$) transition widths are marked by a black cross and the uncertainties are indicated by the red error-bars. (Top) In the case of pure E$2$ transitions, the measured transition widths are reduced by the contribution due to quadrupole deformation, according to Eqs.~\ref{eq2024-1-2}, \ref{eq2024-1-5}, and \ref{eq2024-1-6}, and indicated by dashed arrow, to those which obey the Weisskopf estimate for E$2$ transitions, according to Eq.~\ref{eq2024-1-9}. (Bottom) In the case of mixed M$1+$E$2$ transitions, the measured transition widths are reduced by (i) the Weisskopf estimate for M$1$ transitions, in Eq.~\ref{eq2024-1-7}, indicated by the dotted hollow circle and orange error-bars, and then (ii) by the contribution due to quadrupole deformation, according to Eqs.~\ref{eq2024-1-2}, \ref{eq2024-1-5}, and \ref{eq2024-1-6}, and indicated by dashed arrow, to those which obey the Weisskopf estimate for E$2$ transitions, according to Eq.~\ref{eq2024-1-9}.}
\label{fig2024-1-3}
\end{figure}

In Eq.~\ref{eq2024-1-10}, we know the E$2$ transition lifetimes of the candidate isotope and that of the nearest even-even nuclei, and the Weisskopf estimate contributions to the total E$2$ transition strengths for both, as well as the adopted E$2$ transition strength (and adopted $\beta_2$) for the even-even nuclei. This allowed us to isolate the contribution to the total E$2$ transition strength in the candidate isotope, $B^*_{\beta}(\text{E}2: j_i \rightarrow j_f)$. The empirically measured E$2$ transition width along with the contribution due to the quadrupole deformation, thus extracted have been plotted as a function of $A^{4/15}\mathcal{E}_{\gamma}$ in Figure~\ref{fig2024-1-3}, similar to the Weisskopf estimate for the E$2$ transition lifetime in Figure~\ref{fig2024-1-1}. The Weisskopf estimate for the E$2$ transition lifetime for all candidate isotopes inevitably follows the same slope as in Eq.~\ref{eq2024-1-9}.

The majority of isotopes relevant for EDM searches are odd-even nuclei which have a ground state spin of at least $1/2$ \cite{MohanMurthy2024-dh}. Therefore, when going from the so extracted E$2$ transition strength to the coefficient characterizing the nuclear quadrupole deformation, $\beta_2$, via Eqs.~\ref{eq2024-1-5} and \ref{eq2024-1-6}, we ensured that the right transition matrix elements (in Eq.~\ref{eq2024-1-5}), corresponding to the initial and final spin states involved in the E$2$ transition to the ground state, were used. The transition matrix element used not only the appropriate Clebsch-Gordan coefficients, but also the correction factors arising from relativistic effects and those arising from the interaction of the angular momentum with the spin (see Tables A.1 to A.3 in \emph{Ref.}~\cite{van-Dommelen2012-fg}). These values of $\beta_2$ are listed in Table~\ref{tab2024-1-2}, along with the recommended values for the corresponding E$2$ transition strengths.

\subsection{Quadrupole Deformation from M1+E2 Transitions}

In the case of mixed M$1$+E$2$ transitions, the transition strengths corresponding to the Weisskopf single particle estimates for both the M$1$ and E$2$ transitions, each linked to the emitted photon energy according to Eqs.~\ref{eq2024-1-7} and \ref{eq2024-1-8}, respectively, were calculated for each candidate isotope. Then, like in the case of pure E$2$ transitions, the ratio of M$1$+E$2$ transition lifetime of the candidate isotope to the E$2$ transition lifetime of the nearest even-even nuclei was used,
\begin{eqnarray}
\frac{\tau^{\text{(even-even)}}_{1/2}}{\tau^*_{1/2}} \approx \frac{B^*_{w:\text{E}2}(\mathcal{E}^*_{\gamma})+B^*_{w:\text{M}1}(\mathcal{E}^*_{\gamma})+B^*_{\beta}(\text{E}2: j_i \rightarrow j_f)}{B^{\text{(even-even)}}_{w:\text{E}2}(\mathcal{E}^{\text{(even-even)}}_{\gamma})+B^{\text{(even-even)}}_{\text{adopted}-\beta}(\text{E}2: 0^+ \rightarrow 2^+)},~\label{eq2024-1-11}
\end{eqnarray}
where the lifetime parameter marked with an asterisk corresponds to the candidate isotope's M$1$+E$2$ transition lifetime, and the lifetime parameter marked as (even-even) corresponds to the nearest even-even nuclei's E$2$ transition lifetime. All three Weisskopf estimates, $\{B^*_{w:\text{E}2},B^*_{w:\text{M}1},B^{\text{(even-even)}}_{w:\text{E}2}\}$, the two empirically measured lifetimes, $\{\tau^{\text{(even-even)}}_{1/2},\tau^*_{1/2}\}$, and the adopted E$2$ transition strength due to the quadrupole deformation in the nearest even-even nuclei were known. In this fashion, it was possible to isolate the E$2$ transition strength, $B^*_{\beta}(\text{E}2: j_i \rightarrow j_f)$, corresponding to the nuclear quadrupole deformation.

%errors w.r.t E2
Similar to the case of pure E$2$ transitions, the quadrupole deformation of the candidate isotopes with M$1$+E$2$ transitions was determined using the thus extracted E$2$ transition strength, after carefully considering the two correction factors, and the initial and final spin states. These deformation coefficients extracted have been listed Table~\ref{tab2024-1-3}. In these candidates with mixed M$1$+E$2$ transitions, due to the inclusion of the uncertainty arising from subtracting the contribution to the total width from the M$1$ transition, in addition to that from the E$2$ transition, the thus extracted values of $\beta_2$ are less precise,  compared to those in Table~\ref{tab2024-1-2} obtained from pure E$2$ transitions. In Figure~\ref{fig2024-1-3}, according to Eq.~\ref{eq2024-1-2}, the width corresponding to the Weisskopf estimate for the M$1$ transition was subtracted from the total M$1$+E$2$ transition width. Then, the width correction due to the quadrupole deformation was shown, as a means to bring the M$1$ transition corrected total M$1$+E$2$ transition width down to the width corresponding to the Weisskopf estimate for the E$2$ transition. Naturally, the Weisskopf estimate for E$2$ transition widths scale as the fifth power of transition energy, $\mathcal{E}_{\gamma}$, with the same slope as in Eq.~\ref{eq2024-1-9}.

\pagebreak

\begin{footnotesize}
\centering
\setlength{\tabcolsep}{2pt}
\begin{longtable}[htbp]{l l c c c c c c l}
%\centering
%\begin{tabular}
\caption[]{Table summarizing the theoretical quadrupole deformation coefficient , $\beta^{\text{(FRDM)}}_2$ \cite{Moller2016-gt}, transition energy, $\mathcal{E}_{\gamma}$ \cite{nndc}, recommended E$2$ transition strength, $B^*(\text{E}2)$ and corresponding value of the quadrupole deformation coefficient, $\beta^*_2$, extracted from the E$2$ transition lifetimes. The half-life of the isotope is given as its subscript (where the units are, s: seconds, m: minutes, d: days, and y: years). $^\dagger$: These do not have an uncertainty associated with the lifetime; therefore, one half the value of the transition lifetime was used as its uncertainty, and the uncertainties of the recommended $B^*_{\beta:\text{E}2}$ and $\beta^*_2$ correspond to $68.3\%~$C.I.; and
$^{(k,l)}$: in case the E$2$ transition occurs within or between rotational bands higher than $1$, they are indicated for both the parent and ground state participating in the transition, respectively, within brackets as the superscript. Isotopes marked, gray: have another measurement of the same isotope using the same type of transition; blue: have independent measurement(s) using the other type of transition; yellow: are laser coolable \cite{McClelland2016-od}; red: are ideal MQM candidates \cite{MohanMurthy2024-dh}; orange: are ideal nuclear Schiff moment (NSM) candidates \cite{Mohanmurthy2020-np}; and magenta: are both ideal MQM and NSM candidates.}
\label{tab2024-1-2}\\
\hline
\hline
%\hhline{=========}
$\bm{\text{\bf E}2}$ & $^{A}\cdot_{\tau_{1/2}}$ & $j^{\pi}\rightarrow \bm{j_{\text{\bf Gr.-St.}}^{\pi}}$ & $\mathcal{E}^*_{\gamma}~$(keV) & $\tau^*_{1/2}~$ & & $|\beta|^{\text{(FRDM)}}_2$& $B^*_{\beta:\text{E}2}~(\mathbb{e}^2b^2)$ & $|\beta|^*_2$ \\
\hline
$Z=61$& $^{151}$Pm$^{\dagger(3,1)}_{28.40(4)\text{h}}$ & $1/2^+\rightarrow \bm{5/2^+}$ & $426.451(14)$ & $<200$ & ps & $0.251$ & $>2.8$ & $>0.247$\\%2.8\pm1.9 & 0.247(83)\\
\hline
$Z=62$& $^{151}$Sm$^{(5)}_{90(8)\text{y}}$ & $9/2^- \rightarrow \bm{5/2^-}$ & $294.82(4)$ & $26(7)$ & ps & $0.215$ & $1.36(38)$ & \cellcolor{blue!35}$0.169(24)$\\
\hline
%$Z=63$& $^{152}$Eu$^{(N,1)}_{13.517(9)\text{y}}$ & $1^- \rightarrow \bm{3^-}$ & $65.2969(4)$ & $940(80)$ & ns & $0.224$ & $1.38(15)$ & $0.167(9)$\\
$Z=63$ & $^{152}$Eu$_{13.517(9)\text{y}}$ & $5^- \rightarrow \bm{3^-}$ & $180.6328(4)$ & $2.1(6)$ & ns & $0.224$ & $1.38(41)$ & $0.167(25)$\\
& $^{153}$Eu$_{\infty}$ & $9/2^+ \rightarrow \bm{5/2^+}$ & $193.0654(6)$ & $179.9(91)$ & ps & $0.252$ & $3.52(18)$ & $0.2666(68)$\\
\hline
$Z=64$ & $^{153}$Gd$_{240.4(10)\text{d}}$ & $7/2^- \rightarrow \bm{3/2^-}$ & $93.3429(6)$ & $450(80)$ & ps & $0.215$ & $1.77(31)$ & \cellcolor{blue!35} $0.186(16)$\\
& $^{155}$Gd$_{\infty}$ & $7/2^- \rightarrow \bm{3/2^-}$ & $146.0696(7)$ & $102(11)$ & ps & $0.252$ & $3.88(42)$ & \cellcolor{blue!35} $0.274(15)$\\
& $^{157}$Gd$_{\infty}$ & $7/2^- \rightarrow \bm{3/2^-}$ & $131.451(9)$ & $100(10)$ & ps & $0.271$ & $4.76(47)$ & \cellcolor{blue!35}  $0.303(15)$\\
\hline
$Z=65$ & $^{155}$Tb$^{\dagger}_{5.32(6)\text{d}}$ & $7/2^+ \rightarrow \bm{3/2^+}$ & $155.785(3)$ & $\le200$ & ps & $0.243$ & $\ge3.9$ & \cellcolor{blue!35} $\ge0.270$\\ %$3.9(19)$ & $0.270(67)$\\
& $^{159}$Tb$_{\infty}$ & $7/2^+ \rightarrow \bm{3/2^+}$ & $137.5055(17)$ & $41.3(21)$ & ps & $0.271$ & $5.06(26)$ & \cellcolor{blue!35} $0.3059(78)$\\
%& $^{160}$Tb$^{(4,1)}_{72.3(2)\text{d}}$ & $1^- \rightarrow \bm{3-}$ & $63.6856(2)$ & $60(5)$ & ns & $0.271$ & $5.10(42)$ & $0.306(13)$\\
\hline
$Z=66$ & $^{155}$Dy$_{9.9(2)\text{h}}$ & $7/2^- \rightarrow \bm{3/2^-}$ & $86.767(12)$ & $1.1(2)$ & ns & $0.216$ & $2.37(43)$ & \cellcolor{blue!35} $0.208(19)$\\
& $^{157}$Dy$^{\dagger}_{8.14(4)\text{h}}$ & $7/2^- \rightarrow \bm{3/2^-}$ & $147.724(9)$ & $\le300$ & ps & $0.252$ & $\ge3.8$ & \cellcolor{blue!35} $\ge0.263$\\ %$3.8(19)$ & $0.263(65)$\\
& $^{161}$Dy$_{\infty}$ & $9/2^+ \rightarrow \bm{5/2^+}$ & $100.4033(2)$ & $220(30)$ & ps & $0.272$ & $5.11(70)$ & $0.302(20)$\\
& $^{163}$Dy$_{\infty}$ & $9/2^- \rightarrow \bm{5/2^-}$ & $167.3451(12)$ & $340(60)$ & ps & $0.281$ & $5.40(95)$ & \cellcolor{blue!35} $0.308(28)$\\
\hline
$Z=67$ & \cellcolor{red!50} $^{157}$Ho$_{12.6(2)\text{m}}$ & $11/2^- \rightarrow \bm{7/2^-}$ & $188.07(3)$ & $46(12)$ & ps & $0.235$ & $3.8(10)$ & \cellcolor{blue!35}$0.259(35)$\\
& $^{165}$Ho$_{\infty}$ & \cellcolor{gray!50} $11/2^- \rightarrow \bm{7/2^-}$ & $209.804(11)$ & $12.71(14)$ & ps & $0.293$ & $5.66(32)$ & \cellcolor{blue!35}$0.310(1)$\\
& $^{165}$Ho$^{4,1}_{\infty}$ & \cellcolor{gray!50} $3/2^- \rightarrow \bm{7/2^-}$ & $515.476(11)$ & $10.4(11)$ & ps & $0.293$ & $5.67(57)$ & \cellcolor{blue!35} $0.310(16)$\\
\hline
\cellcolor{yellow!50} $Z=68$ & $^{159}$Er$^{\dagger}_{36(1)\text{m}}$ & $7/2^- \rightarrow \bm{3/2^-}$ & $144.232(14)$ & $\le170$ & ps & $0.235$ & $\ge3.3$ & \cellcolor{blue!35} $\ge0.237$\\ %$3.3(17)$ & $0.237(61)$\\
\cellcolor{yellow!50} & $^{161}$Er$^{\dagger}_{3.21(3)\text{h}}$ & $7/2^- \rightarrow \bm{3/2^-}$ & $143.89(3)$ & $<180$ & ps & $0.263$ & $\ge4.3$ & $\ge0.263$\\%$4.3(22)$ & $0.263(69)$\\
\cellcolor{yellow!50} &\cellcolor{red!50}$^{165}$Er$^{(4,2)}_{10.36(4)\text{h}}$ & $1/2^- \rightarrow \bm{5/2^-}$ & $297.371(4)$ & $700(80)$ & ps & $0.282$ & $5.5(10)$ & $0.302(27)$\\
%& $^{167}$Er$^{(4,N)}_{\infty}$ & $3/2^+ \rightarrow \bm{7/2^+}$ & $531.54(4)$ & $19.3(23)$ & ps & $0.294$ & $5.94(83)$ & $0.311(22)$\\
%& $^{167}$Er$^{(N)}_{\infty}$ & $(11/2)^+ \rightarrow \bm{7/2^+}$ & $177.971(7)$ & $55(6)$ & ps & $0.294$ & $6.0(27)$ & $0.313(70)$\\
\hline
\cellcolor{yellow!50} $Z=69$ & $^{163}$Tm$^{(8,2)}_{1.810(5)\text{h}}$ & $5/2^+ \rightarrow \bm{1/2^+}$ & $136.71(2)$ & $60(10)$ & ps & $0.274$ & $5.06(85)$ & \cellcolor{blue!35} $0.286(24)$\\
\cellcolor{yellow!50} & $^{167}$Tm$_{9.25(2)\text{d}}$ & $5/2^+ \rightarrow \bm{1/2^+}$ & $116.575(18)$ & $66(7)$ & ps & $0.284$ & $5.95(64)$ & \cellcolor{blue!35} $0.308(16)$\\
\cellcolor{yellow!50} & $^{169}$Tm$_{\infty}$ & $5/2^+ \rightarrow \bm{1/2^+}$ & $118.18945(11)$ & $62(3)$ & ps & $0.295$ & $5.91(28)$ & \cellcolor{blue!35} $0.3051(72)$\\
\cellcolor{yellow!50} & $^{170}$Tm$_{128.6(3)\text{d}}$ & $3^- \rightarrow \bm{1^-}$ & $118.18945(11)$ & $600(30)$ & ps & $0.295$ & $5.96(30)$ & $0.3059(77)$\\
\cellcolor{yellow!50} & $^{171}$Tm$_{1.92(1)\text{y}}$ & $5/2^+ \rightarrow \bm{1/2^+}$ & $116.657(1)$ & $55(13)$ & ps & $0.296$ & $5.9(14)$ & $0.303(36)$\\
\hline
\cellcolor{yellow!50} $Z=70$ & $^{165}$Yb$_{9.8(5)\text{m}}$ & $9/2^- \rightarrow \bm{5/2^-}$ & $197.55(13)$ & $179(5)$ & ps & $0.273$ & $4.37(13)$ & $0.2608(38)$\\
\cellcolor{yellow!50} & \cellcolor{red!50}$^{167}$Yb$^{\dagger}_{17.5(2)\text{m}}$ & $9/2^- \rightarrow \bm{5/2^-}$ & $178.857(13)$ & $\le230$ & ps & $0.274$ & $\ge5.2$ & $\ge0.283$\\%$5.2(26)$ & $0.283(71)$\\
\cellcolor{yellow!50} & $^{171}$Yb$_{\infty}$ & $5/2^- \rightarrow \bm{1/2^-}$ & $75.882(2)$ & $1.64(16)$ & ns & $0.295$ & $5.67(56)$ & \cellcolor{blue!35} $0.293(14)$\\
\cellcolor{yellow!50}& $^{173}$Yb$_{\infty}$ & $9/2^- \rightarrow \bm{5/2^-}$ & $179.364(9)$ & $32(4)$ & ps & $0.297$ & $6.07(76)$ & \cellcolor{blue!35} $0.303(18)$\\
\hline
$Z=71$ & $^{175}$Lu$_{\infty}$ & $11/2^+ \rightarrow \bm{7/2^+}$ & $251.465(7)$ & $32.4(16)$ & ps & $0.287$ & $5.84(30)$ & \cellcolor{blue!35} $0.2914(76)$\\
& $^{176}$Lu$^{(2,2)}_{3.76(7)\times10^{10}\text{y}}$ & $9^- \rightarrow \bm{7-}$ & $388.877(4)$ & $7.5(11)$ & ps & $0.278$ & $7.6(11)$ & $0.292(21)$\\
\hline
%$Z=72$ & $^{173}$Hf$^{(2,1)}_{23.6(1)\text{h}}$ & $5/2^- \rightarrow \bm{1/2^-}$ & $107.16(5)$ & $180(8)$ & ns & $0.295$ & $4.37(83)$ & $0.249(24)$\\
$Z=72$ & $^{177}$Hf$_{\infty}$ & $11/2^- \rightarrow \bm{7/2^-}$ & $249.6744(4)$ & $107(10)$ & ps & $0.277$ & $5.41(51)$ & $0.275(13)$\\
& $^{179}$Hf$^{(2)}_{\infty}$ & $13/2^+ \rightarrow \bm{9/2^+}$ & $268.92(6)$ & $21(3)$ & ps & $0.278$ & $4.76(68)$ & \cellcolor{blue!35} $0.257(18)$\\
\hline
$Z=73$ & $^{181}$Ta$^{(5)}_{\infty}$ & $11/2^+ \rightarrow \bm{7/2^+}$ & $301.622(22)$ & $16(3)$ & ps & $0.269$ & $4.69(89)$ & \cellcolor{blue!35} $0.251(24)$\\
\hline
%$Z=74$ & $^{177}$W$^{(2,1)}_{132.4(20)\text{m}}$ & $5/2^- \rightarrow \bm{1/2^-}$ & $101.23(8)$ & $38(8)$ & ns & $0.276$ & $5.0(11)$ & $0.257(29)$\\
$Z=74$ & $^{183}$W$_{\ge6.7\times10^{20}\text{y}}$ & $5/2^- \rightarrow \bm{1/2^-}$ & $99.0791(9)$ & $726(20)$ & ps & $0.250$ & $4.20(12)$ & $0.2337(34)$\\
& $^{185}$W$^{\dagger}_{75.1(3)\text{d}}$ & $7/2^- \rightarrow \bm{3/2^-}$ & $173.702(16)$ & $<1.5$ & ns & $0.241$ & $>3.7$ & $>0.219$\\ %$3.7(20)$ & $0.219(59)$\\
\hline
$Z=75$ & $^{185}$Re$_{\infty}$ & \cellcolor{gray!50} $9/2^+ \rightarrow \bm{5/2^+}$ & $284.2(3)$ & $5.6(15)$ & ps & $0.221$ & $3.7(10)$ & $0.214(30)$\\
& $^{185}$Re$^{(7,1)}_{\infty}$ & \cellcolor{gray!50} $1/2^+ \rightarrow \bm{5/2^+}$ & $646.134(4)$ & $6.3(4)$ & ps & $0.221$ & $3.75(31)$ & $0.217(9)$\\
%& $^{186}$Re$^{(7,4)}_{3.7185(5)\text{d}}$ & $3^- \rightarrow \bm{1^-}$ & $99.361(3)$ & $25.5(25)$ & ns & $0.221$ & $3.79(39)$ & $0.218(11)$\\
& $^{187}$Re$_{4.33(7)\times10^{10}\text{y}}$ & $9/2^+ \rightarrow \bm{5/2^+}$ & $303.36(7)$ & $5.2(18)$ & ps & $0.212$ & $3.6(12)$ & \cellcolor{blue!35} $0.212(35)$\\
%& $^{187}$Re$^{(N,4)}_{4.33(7)\times10^{10}\text{y}}$ & $1/2^+ \rightarrow \bm{5/2^+}$ & $303.36(7)$ & $13(3)$ & ps & $0.212$ & $3.56(85)$ & $0.210(26)$\\
\hline
%$Z=76$ & $^{177}$Os$^{(2,1)}_{3.0(2)\text{m}}$ & $5/2^- \rightarrow \bm{1/2^-}$ & $152.30(24)$ & $40(3)$ & ns & $0.255$ & $3.4(10)$ & $0.207(31)$\\
$Z=76$ & $^{187}$Os$^{(2,1)}_{\infty}$ & $5/2^- \rightarrow \bm{1/2^-}$ & $75.016(22)$ & $2.16(16)$ & ns & $0.212$ & $2.97(23)$ & \cellcolor{blue!35} $0.1900(74)$\\
& $^{189}$Os$^{(4,2)}_{\infty}$ & \cellcolor{gray!35} $7/2^- \rightarrow \bm{3/2^-}$ & $216.67(2)$ & $410(20)$ & ps & $0.183$ & $2.40(12)$ & \cellcolor{blue!35} $0.1715(40)$\\
& $^{189}$Os$^{(2)}_{\infty}$ & \cellcolor{gray!35} $7/2^- \rightarrow \bm{3/2^-}$ & $219.39(2)$ & $260(11)$ & ps & $0.183$ & $2.50(10)$ & \cellcolor{blue!35} $0.1735(35)$\\
\hline
%& $^{185}$Ir$^{(3)}_{14.4(1)\text{h}}$ & \cellcolor{gray!50} $9/2^- \rightarrow \bm{5/2^-}$ & $5.8(1)$ & $5(1)$ & ns & $0.211$ & $3.11(68)$ & $0.193(21)$\\
$Z=77$ & $^{185}$Ir$^{(3)}_{14.4(1)\text{h}}$ & %\cellcolor{gray!50}
$1/2^- \rightarrow \bm{5/2^-}$ & $135.3(1)$ & $290(30)$ & ps & $0.211$ & $3.11(31)$ & $0.194(10)$\\
& $^{189}$Ir$^{\dagger(2)}_{13.2(1)\text{d}}$ & $7/2^+ \rightarrow \bm{3/2^+}$ & $300.50(4)$ & $<20$ & ps & $0.164$ & $>2.5$ & \cellcolor{blue!35} $>0.170 $ \\ %$2.5(13)$ & $0.170(44)$\\
%& $^{191}$Ir$_{\infty}$ & $7/2^+ \rightarrow \bm{3/2^+}$ & $686.1(1)$ & $2.7(3)$ & ps & $0.155$ & $2.38(28)$ & $0.167(10)$\\
& $^{191}$Ir$_{\infty}$ & $7/2^+ \rightarrow \bm{3/2^+}$ & $343.32(4)$ & $20.4(8)$ & ps & $0.155$ & $2.38(10)$ & \cellcolor{blue!35} $0.167(4)$\\
& $^{193}$Ir$_{\infty}$ & $7/2^+ \rightarrow \bm{3/2^+}$ & $357.768(5)$ & $18.7(7)$ & ps & $0.136$ & $2.055(79)$ & \cellcolor{blue!35} $0.1543(30)$\\
\hline
$Z=78$ & $^{195}$Pt$^{(3,1)}_{\infty}$ & \cellcolor{gray!50} $5/2^- \rightarrow \bm{1/2^-}$ & $129.772(3)$ & $670(30)$ & ps & $0.148$ & $1.661(70)$ & \cellcolor{blue!35} $0.1365(29)$\\
& $^{195}$Pt$_{\infty}$ & \cellcolor{gray!50} $5/2^- \rightarrow \bm{1/2^-}$ & $239.264(4)$ & $70(9)$ & ps & $0.148$ & $1.67(20)$ & \cellcolor{blue!35} $0.1370(83)$\\
%& $^{197}$Pt$^{(N)}_{19.8915(19)\text{h}}$ & $5/2^- \rightarrow \bm{1/2^-}$ & $53.088(19)$ & $16.58(17)$ & ns & $0.148$ & $1.382(14)$ & $0.12406(64)$\\
\hline
%$Z=79$ & $^{185}$Au$^{(N,4)}_{4.25(6)\text{m}}$ & $(9/2)^- \rightarrow \bm{5/2^-}$ & $8.9(1)$ & $4.8(4)$ & ns & $0.156$ & $3.98(40)$ & $0.212(11)$\\
%& $^{196}$Au$^{(N,N')}_{6.1669(6)\text{d}}$ & $4^- \rightarrow \bm{2^-}$ & $212.80(4)$ & $1.79(15)$ & ns & $0.139$ & $1.67(18)$ & $0.1349(74)$\\
%$Z=80$ & $^{197}$Au$^{(N)}_{\infty}$ & $7/2^+ \rightarrow \bm{3/2^+}$ & $547.5(3)$ & $4.61(16)$ & ps & $0.131$ & $1.383(52)$ & $0.1225(23)$\\
%$Z=80$ & $^{198}$Au$^{(N)}_{2.6941(2)\text{2}}$ & $4^- \rightarrow \bm{2^-}$ & $214.9720(9)$ & $0.4(2)$ & ps & $0.131$ & $1.39(74)$ & $0.122(33)$\\
%\hline
%$Z=80$ & $^{195}$Hg$^{(N)}_{10.53(3)\text{h}}$ & $5/2^- \rightarrow \bm{1/2^-}$ & $53.289(20)$ & $720(30)$ & ps & $0.130$ & $1.030(43)$ & $0.1048(22)$\\
% & $^{197}$Hg$^{(N)}_{64.14(5)\text{h}}$ & $5/2^- \rightarrow \bm{1/2^-}$ & $133.96(4)$ & $8.07(16)$ & ns & $0.130$ & $1.133(35)$ & $0.1095(17)$\\
% & $^{199}$Hg$^{(N)}_{\infty}$ & $5/2^- \rightarrow \bm{1/2^-}$ & $158.3790(1)$ & $2.45(3)$ & ns & $0.122$ & $0.994(20)$ & $0.1022(10)$\\
% & $^{201}$Hg$^{(N,N')}_{\infty}$ & $7/2^- \rightarrow \bm{3/2^-}$ & $414.543(18)$ & $21.0(9)$ & ps & $0.105$ & $1.133(35)$ & $0.0947(22)$\\
%\hline
%HERE!!! Remove al Hg isotopes from paper
%$Z=81$ & $^{200}$Tl$^{(N)}_{26.1(1)\text{h}}$ & $0^- \rightarrow \bm{2^-}$ & $147.634(21)$ & $7.10(15)$ & ns & $0.044$ & $1.00(4)$ & $0.1011(20)$\\
%\hline
 \cellcolor{yellow!50} $Z=87$ & \cellcolor{orange!50} $^{227}$Fr$^{(3,1)}_{2.47(3)\text{m}}$ & $5/2^+ \rightarrow \bm{1/2^+}$ & $164.95(4)$ & $49(8)$ & ps & $0.181$ & $2.37(40)$ & \cellcolor{blue!35} $0.139(12)$\\
\hline
\cellcolor{yellow!50}  $Z=88$ & \cellcolor{magenta!50} $^{223}$Ra$^{\dagger}_{11.43(5)\text{d}}$ & $7/2^+ \rightarrow \bm{3/2^+}$ & $61.424(10)$ & $\simeq 600$ & ps & $0.156$ & $\simeq 4.5$ & $\simeq 0.19$\\%$4.5(8)$ & $0.19(4)$\\
\cellcolor{yellow!50}  & \cellcolor{orange!50} $^{225}$Ra$_{14.9(2)\text{d}}$ & $5/2^+ \rightarrow \bm{1/2^+}$ & $25.41(2)$ & $880(40)$ & ps & $0.164$ & $4.02(18)$ & $0.1794(39)$\\
%\cellcolor{yellow!50}  & \cellcolor{red!50} $^{229}$Ra$^{(N,3)}_{4.0(2)\text{m}}$ & $1/2^+ \rightarrow \bm{5/2^+}$ & $142.67(6)$ & $17.23(12)$ & ns & $0.189$ & $5.93(53)$ & $0.217(10)$\\
\hline
$Z=90$ & \cellcolor{magenta!50} $^{229}$Th$_{7880(120)\text{y}}$ & $9/2^+ \rightarrow \bm{5/2^+}$ & $97.13595(24)$ & $147(12)$ & ps & $0.190$ & $6.95(58)$ & \cellcolor{blue!35} $0.229(1)$\\
& \cellcolor{red!50} $^{231}$Th$^{\dagger(6,1)}_{25.57(8)\text{h}}$ & $1/2^+ \rightarrow \bm{5/2^+}$ & $247.5867(20)$ & $<74$ & ps & $0.198$ & $>8.2$ & $>0.249$\\ %& $8.2(42)$ & $0.249(64)$\\
\hline
$Z=91$ & \cellcolor{red!50} $^{231}$Pa$_{32570(130)\text{y}}$ & $7/2^- \rightarrow \bm{3/2^-}$ & $58.5696(24)$ & $274(10)$ & ps & $0.198$ & $8.24(31)$ & $0.2462(47)$\\
\hline
$Z=92$ & $^{233}$U$^{(5,1)}_{1.5919(15)\times10^{5}\text{y}}$ & $1/2^+ \rightarrow \bm{5/2^+}$ & $398.496(8)$ & $55(20)$ & ps & $0.207$ & $10.3(41)$ & \cellcolor{blue!35} $0.272(53)$\\
& \cellcolor{red!50} $^{235}$U$^{(1,2)}_{7.04(1)\times10^{8}\text{y}}$ & $11/2^- \rightarrow \bm{7/2^-}$ & $103.903(8)$ & $33(5)$ & ps & $0.215$ & $10.2(15)$ & \cellcolor{blue!35} $0.270(53)$\\
\hline
$Z=94$ & $^{239}$Pu$_{24110(30)\text{y}}$ & $5/2^+ \rightarrow \bm{1/2^+}$ & $57.275(2)$ & $101(5)$ & ps & $0.223$ & $12.55(63)$ & $0.2909(72)$\\
\hline
$Z=97$ & \cellcolor{red!50} $^{249}$Bk$_{327.2(3)\text{d}}$ & $11/2^+ \rightarrow \bm{7/2^+}$ & $93.759(8)$ & $5(1)$ & ps & $0.235$ & $15.1(29)$ & \cellcolor{blue!35} $0.305(30)$\\
%\hhline{=========}
\hline
\hline
\end{longtable}
%\end{table}
\end{footnotesize}

\begin{footnotesize}
\centering
\setlength{\tabcolsep}{2pt}
\begin{longtable}[htbp]{l l c c c c c c l}
%\centering
%\begin{tabular}
\caption[]{Table summarizing the theoretical quadrupole deformation coefficient , $\beta^{\text{(FRDM)}}_2$ \cite{Moller2016-gt}, transition energy, $\mathcal{E}_{\gamma}$ \cite{nndc}, recommended E$2$ transition strength, $B^*(\text{E}2)$ and corresponding value of the quadrupole deformation coefficient, $\beta^*_2$, extracted from the M$1+$E$2$ transition lifetimes. It is important to note that the transition considered here is the M$1+$E$2$ transition, but the recommended $B^*(\text{E}2)$ is the pure E$2$ transition equivalent for the mixed M$1+$E$2$ transition. The half-life of the isotope is given as its subscript and the color codes refer to the same as in Table~\ref{tab2024-1-2}.}
\label{tab2024-1-3}\\
\hline
\hline
%\hhline{=========}
$\bm{\text{\bf M}1+\text{\bf E}2}$ & $^{A}\cdot_{\tau_{1/2}}$ & $j^{\pi}\rightarrow \bm{j_{\text{\bf Gr.-St.}}^{\pi}}$ & $\mathcal{E}^*_{\gamma}~$(keV) & $\tau^*_{1/2}~$ & & $|\beta|^{\text{(FRDM)}}_2$& $B^*_{\beta:\text{E}2}~(\mathbb{e}^2b^2)$ & $|\beta|^*_2$ \\
\hline
%$Z=62$& $^{151}$Sm$^{(11,5)}_{90(8)\text{y}}$ & \cellcolor{gray!50} $3/2^- \rightarrow \bm{5/2^-}$ & $4.821(3)$ & $35(1)$ & ns & $0.215$ & $1.364(39)$ & \cellcolor{blue!35} $0.1694(24)$\\
$Z=62$ & $^{151}$Sm$^{(4,5)}_{90(8)\text{y}}$ & \cellcolor{gray!50} $7/2^- \rightarrow \bm{5/2^-}$ & $65.826(5)$ & $400(60)$ & ps & $0.215$ & $1.36(29)$ & \cellcolor{blue!35} $0.169(18)$\\
& $^{151}$Sm$^{(4,5)}_{90(8)\text{y}}$ & \cellcolor{gray!50} $3/2^- \rightarrow \bm{5/2^-}$ & $104.831(5)$ & $480(30)$ & ps & $0.215$ & $1.37(13)$ & \cellcolor{blue!35} $0.1704(75)$\\
\hline
%$Z=63$& $^{151}$Eu$^{(1,N)}_{>1.7\times10^{18}\text{y}}$ & $7/2^+ \rightarrow \bm{5/2^+}$ & $21.541(3)$ & $9.6(3)$ & ns & $0.215$ & $1.363(69)$ & $0.1667(42)$\\
%\hline
$Z=64$ & $^{153}$Gd$_{240.4(10)\text{d}}$ & \cellcolor{gray!50} $5/2^- \rightarrow \bm{3/2^-}$ & $41.5568(4)$ & $4.08(6)$ & ns & $0.215$ & $1.77(31)$ & \cellcolor{blue!35} $0.186(20)$\\
& $^{153}$Gd$^{(3,1)}_{240.4(10)\text{d}}$ & \cellcolor{gray!50} $(5/2)^- \rightarrow \bm{3/2^-}$ & $109.7560(7)$ & $243(14)$ & ps & $0.215$ & $1.76(65)$ & \cellcolor{blue!35}$0.184(35)$\\
& $^{155}$Gd$^{(2,1)}_{\infty}$ & $5/2^- \rightarrow \bm{3/2^-}$ & $60.0106(6)$ & $196(15)$ & ps & $0.252$ & $3.89(33)$ & \cellcolor{blue!35} $0.275(12)$\\
& $^{157}$Gd$_{\infty}$ & $5/2^- \rightarrow \bm{3/2^-}$ & $54.536(9)$ & $130(8)$ & ps & $0.271$ & $4.75(32)$ & \cellcolor{blue!35} $0.301(10)$\\
\hline
%$Z=65$ & $^{151}$Tb$_{17.609(14)\text{h}}$ & $(5/2^+) \rightarrow \bm{1/2^{(+)}}$ & $72.39(3)$ & $920(30)$ & ps & $0.161$ & $0.24(1)$ & $0.068(14)$\\
$Z=65$ & $^{153}$Tb$^{(4,1)}_{2.34(1)\text{d}}$ & $7/2^+ \rightarrow \bm{5/2^+}$ & $80.720(2)$ & $490(20)$ & ps & $0.216$ & $0.24(1)$ & $0.068(14)$\\
& $^{155}$Tb$_{5.32(6)\text{d}}$ & $5/2^+ \rightarrow \bm{3/2^+}$ & $65.4622(24)$ & $250(30)$ & ps & $0.243$ & $3.90(56)$ & \cellcolor{blue!35} $0.27(2)$\\
& $^{157}$Tb$^{(2,1)}_{71(7)\text{y}}$ & $5/2^+ \rightarrow \bm{3/2^+}$ & $60.881(3)$ & $490(120)$ & ps & $0.271$ & $4.8(16)$ & $0.30(5)$\\
 & $^{159}$Tb$_{\infty}$ & \cellcolor{gray!50} $5/2^+ \rightarrow \bm{3/2^+}$ & $57.9964(15)$ & $55.0(22)$ & ps & $0.271$ & $5.07(19)$ & \cellcolor{blue!35} $0.3063(57)$\\
& $^{159}$Tb$^{(5,1)}_{\infty}$ & \cellcolor{gray!50} $1/2^+ \rightarrow \bm{3/2^+}$ & $580.808(6)$ & $0.76(10)$ & ps & $0.271$ & $5.1(10)$ & \cellcolor{blue!35} $0.307(29)$\\
\hline
$Z=66$ & $^{155}$Dy$^{(2,1)}_{9.9(2)\text{h}}$ & $5/2^- \rightarrow \bm{3/2^-}$ & $39.384(9)$ & $3.34(3)$ & ns & $0.216$ & $2.37(42)$ & \cellcolor{blue!35} $0.208(10)$\\
& $^{157}$Dy$^{\dagger(2,1)}_{8.14(4)\text{h}}$ & $5/2^- \rightarrow \bm{3/2^-}$ & $61.141(13)$ & $300$ & ps & $0.252$ & $3.8$ & \cellcolor{blue!35} $0.263$\\ %$3.8(23)$ & $0.263(8)$\\
& $^{159}$Dy$^{(2,1)}_{145.3(14)\text{d}}$ & $5/2^- \rightarrow \bm{3/2^-}$ & $56.626(6)$ & $210(40)$ & ps & $0.271$ & $4.7(10)$ & $0.290(31)$\\
%& $^{161}$Dy$^{(N,1)}_{\infty}$ & $7/2^+ \rightarrow \bm{5/2^+}$ & $43.8201(7)$ & $830(60)$ & ps & $0.271$ & $5.10(46)$ & $0.301(14)$\\
& $^{163}$Dy$^{(2,1)}_{\infty}$ & $7/2^- \rightarrow \bm{5/2^-}$ & $73.4448(4)$ & $1.51(5)$ & ns & $0.281$ & $5.47(24)$ & \cellcolor{blue!35} $0.310(7)$\\
& $^{165}$Dy$^{\dagger}_{2.331(4)\text{h}}$ & $(9/2)^+ \rightarrow \bm{7/2^+}$ & $83.3942(15)$ & $<35$ & ps & $0.293$ & $>5.7$ & $>0.316$\\ %& $5.7(31)$ & $0.316(85)$\\
\hline
$Z=67$ & \cellcolor{red!50} $^{157}$Ho$^{\dagger(2,1)}_{12.6(2)\text{m}}$ & $9/2^- \rightarrow \bm{7/2^-}$ & $83.58(3)$ & $\le300$ & ps & $0.235$ & $>3.8$ & \cellcolor{blue!35}$>0.26$\\  %$3.8(31)$ & $0.26(10)$\\
& $^{165}$Ho$_{\infty}$ & $9/2^- \rightarrow \bm{7/2^-}$ & $94.700(3)$ & $22.2(8)$ & ps & $0.293$ & $5.67(26)$ & \cellcolor{blue!35} $0.3103(71)$\\
\hline
\cellcolor{yellow!50} $Z=68$ & $^{159}$Er$^{\dagger(2,1)}_{36(1)\text{m}}$ & $5/2^- \rightarrow \bm{3/2^-}$ & $59.249(14)$ & $\le300$ & ps & $0.235$ & $\ge3.3$ & \cellcolor{blue!35} $\ge0.237$\\ %$3.3(19)$ & $0.237(68)$\\
%& $^{161}$Er$^{\dagger(N,1)}_{3.21(3)\text{h}}$ & $5/2^- \rightarrow \bm{3/2^-}$ & $59.501(24)$ & $\le150$ & ps & $0.263$ & $\ge4.4$ & $\ge0.272$\\ %$4.4(24)$ & $0.272(74)$\\
%& $^{167}$Er$^{(1,N)}_{\infty}$ & $9/2^+ \rightarrow \bm{7/2^+}$ & $79.3221(13)$ & $119(9)$ & ps & $0.294$ & $5.93(52)$ & $0.311(14)$\\
\hline
\cellcolor{yellow!50} $Z=69$ & $^{163}$Tm$^{\dagger(1,2)}_{1.810(5)\text{h}}$ & $3/2^+ \rightarrow \bm{1/2^+}$ & $13.52(2)$ & $<900$ & ps & $0.274$ & $>5.1$ & \cellcolor{blue!35} $>0.287$\\ %& $5.1(25)$ & $0.286(70)$\\
\cellcolor{yellow!50} & $^{165}$Tm$^{(6,5)}_{30.06(3)\text{h}}$ & $3/2^+ \rightarrow \bm{1/2^+}$ & $11.51(5)$ & $750(50)$ & ps & $0.274$ & $5.53(39)$ & $0.298(10)$\\
\cellcolor{yellow!50} & $^{167}$Tm$_{9.25(2)\text{d}}$ & $3/2^+ \rightarrow \bm{1/2^+}$ & $10.417(21)$ & $950(50)$ & ps & $0.284$ & $5.93(32)$ & \cellcolor{blue!35} $0.3069(83)$\\
\cellcolor{yellow!50} & $^{169}$Tm$_{\infty}$ & $3/2^+ \rightarrow \bm{1/2^+}$ & $8.41017(11)$ & $4.09(5)$ & ns & $0.295$ & $5.910(74)$ & \cellcolor{blue!35}$0.305(2)$\\
%\cellcolor{yellow!50}& $^{171}$Tm$_{1.92(1)\text{y}}$ & $3/2^+ \rightarrow \bm{1/2^+}$ & $5.0359(11)$ & $4.77(8)$ & ns & $0.296$ & $5.87(10)$ & $0.300(26)$\\
\hline
\cellcolor{yellow!50} $Z=70$ & $^{171}$Yb$_{\infty}$ & $3/2^- \rightarrow \bm{1/2^-}$ & $66.732(2)$ & $790(50)$ & ps & $0.295$ & $5.7(10)$ & \cellcolor{blue!35} $0.295(25)$\\
\cellcolor{yellow!50} & $^{173}$Yb$_{\infty}$ & $7/2^- \rightarrow \bm{5/2^-}$ & $78.647(12)$ & $46(5)$ & ps & $0.297$ & $6.09(73)$ & \cellcolor{blue!35} $0.310(17)$\\
\hline
$Z=71$ & $^{175}$Lu$_{\infty}$ & $9/2^+ \rightarrow \bm{7/2^+}$ & $113.806(4)$ & $90(4)$ & ps & $0.287$ & $5.80(46)$ & \cellcolor{blue!35} $0.290(11)$\\
& $^{177}$Lu$_{6.6443(9)\text{d}}$ & $9/2^+ \rightarrow \bm{7/2^+}$ & $121.6214(4)$ & $117(4)$ & ps & $0.278$ & $5.40(67)$ & $0.279(18)$\\
\hline
$Z=72$ & $^{179}$Hf$^{(2)}_{\infty}$ & $11/2^+ \rightarrow \bm{9/2^+}$ & $122.7904(24)$ & $37(3)$ & ps & $0.278$ & $4.81(47)$ & \cellcolor{blue!35} $0.259(13)$\\
\hline
$Z=73$ & $^{181}$Ta$^{(6,5)}_{\infty}$ & $9/2^+ \rightarrow \bm{7/2^+}$ & $136.262(13)$ & $39.5(16)$ & ps & $0.269$ & $4.86(30)$ & \cellcolor{blue!35} $0.2557(78)$\\
\hline
%$Z=74$ & $^{183}$W$^{(N,1)}_{\ge6.7\times10^{20}\text{y}}$ & $3/2^- \rightarrow \bm{1/2^-}$ & $46.4838(5)$ & $185(4)$ & ps & $0.250$ & $4.21(10)$ & $0.2340(28)$\\
%\hline
%$Z=75$ & $^{185}$Re$^{(N,1)}_{\infty}$ & $7/2^+ \rightarrow \bm{5/2^+}$ & $125.3587(9)$ & $10.2(15)$ & ps & $0.221$ & $3.77(58)$ & $0.217(28)$\\
$Z=75$ & $^{187}$Re$_{4.33(7)\times10^{10}\text{y}}$ & $7/2^+ \rightarrow \bm{5/2^+}$ & $134.244(4)$ & $10.6(7)$ & ps & $0.212$ & $3.57(26)$ & \cellcolor{blue!35} $0.2110(76)$\\
\hline
$Z=76$ & $^{187}$Os$^{(2,1)}_{\infty}$ & $3/2^- \rightarrow \bm{1/2^-}$ & $9.756(19)$ & $2.38(18)$ & ns & $0.212$ & $2.99(24)$ & \cellcolor{blue!35} $0.1910(80)$\\
 & $^{189}$Os$^{(1,2)}_{\infty}$ & $7/2^- \rightarrow \bm{3/2^-}$ & $36.20(2)$ & $520(30)$ & ps & $0.183$ & $2.50(15)$ & \cellcolor{blue!35} $0.1735(52)$\\
 \hline
$Z=77$ & $^{187}$Ir$_{10.5(3)\text{h}}$ & $5/2^+ \rightarrow \bm{3/2^+}$ & $110.074(22)$ & $120(15)$ & ps & $0.183$ & $2.97(62)$ & $0.187(19)$\\
& $^{189}$Ir$^{(1,2)}_{13.2(1)\text{d}}$ & $5/2^+ \rightarrow \bm{3/2^+}$ & $113.831(23)$ & $76(18)$ & ps & $0.164$ & $2.5(8)$ & \cellcolor{blue!35} $0.171(27)$\\
%& $^{191}$Ir$^{(2,1)}_{\infty}$ & $1/2^+ \rightarrow \bm{3/2^+}$ & $82.4241(23)$ & $4.10(7)$ & ns & $0.155$ & $2.38(20)$ & $0.1667(69)$\\
& $^{191}$Ir$_{\infty}$ & $5/2^+ \rightarrow \bm{3/2^+}$ & $129.426(3)$ & $89.7(12)$ & ps & $0.155$ & $2.38(20)$ & \cellcolor{blue!35} $0.1667(69)$\\
& $^{193}$Ir$_{\infty}$ & $5/2^+ \rightarrow \bm{3/2^+}$ & $138.941(5)$ & $69.7(10)$ & ps & $0.136$ & $2.06(17)$ & \cellcolor{blue!35} $0.1546(65)$\\
\hline
%$Z=78$ & $^{187}$Pt$^{(6,5)}_{2.35(3)\text{h}}$ & $(5/2^-) \rightarrow \bm{3/2^-}$ & $25.53(11)$ & $700(100)$ & ps & $0.229$ & $3.24(49)$ & $0.193(14)$\\
$Z=78$ & $^{195}$Pt$^{(2,1)}_{\infty}$ & $3/2^- \rightarrow \bm{1/2^-}$ & $129.772(3)$ & $170(19)$ & ps & $0.148$ & $1.661(70)$ & \cellcolor{blue!35} $0.1365(29)$\\
\hline
%$Z=79$ & $^{185}$Au$^{(N,4)}_{4.25(6)\text{m}}$ & $(3/2)^- \rightarrow \bm{5/2^-}$ & $35.78(5)$ & $540(50)$ & ps & $0.156$ & $3.97(39)$ & $0.212(10)$\\
$Z=79$ & $^{193}$Au$_{17.65(15)\text{h}}$ & $5/2^+ \rightarrow \bm{3/2^+}$ & $257.986(21)$ & $45(20)$ & ps & $0.139$ & $1.90(94)$ & $0.145(45)$\\
 & $^{199}$Au$_{3.139(7)\text{d}}$ & $1/2^+ \rightarrow \bm{3/2^+}$ & $77.1942(19)$ & $1.3(2)$ & ns & $0.131$ & $1.57(13)$ & $0.120(7)$\\
\hline
%$Z=80$ & $^{195}$Hg$^{\dagger(N)}_{10.53(3)\text{h}}$ & $3/2^- \rightarrow \bm{1/2^-}$ & $37.083(19)$ & $<50$ & ps & $0.130$ & $>1.05$ & $>0.106$\\ %$1.05(54)$ & $0.106(27)$\\
%$Z=80$ & $^{201}$Hg$^{(N)}_{\infty}$ & $5/2^- \rightarrow \bm{3/2^-}$ & $26.2738(3)$ & $629(18)$ & ps & $0.105$ & $0.859(53)$ & $0.0947(14)$\\
%\hline
%$Z=86$ & $^{219}$Rn$^{(N)}_{3.96(1)\text{s}}$ & $(7/2)^+ \rightarrow \bm{5/2^+}$ & $14.400(10)$ & $875(30)$ & ps & $0.103$ & $0.902(77)$ & $0.0877(38)$\\
%& $^{219}$Rn$^{(N)}_{3.96(1)\text{s}}$ & $3/2^+ \rightarrow \bm{5/2^+}$ & $269.475(8)$ & $14.2(23)$ & ps & $0.103$ & $0.92(44)$ & $0.089(21)$\\
%\hline
\cellcolor{yellow!50} $Z=87$ & \cellcolor{orange!50} $^{221}$Fr$^{(3,1)}_{4.9(2)\text{m}}$ & \cellcolor{gray!50} $(3/2)^- \rightarrow \bm{5/2^-}$ & $36.64(4)$ & $1.5(2)$ & ns & $0.120$ & $1.89(30)$ & $0.125(10)$\\
\cellcolor{yellow!50} & \cellcolor{orange!50}$^{221}$Fr$_{4.9(2)\text{m}}$ & \cellcolor{gray!50} $(3/2)^- \rightarrow \bm{5/2^-}$ & $99.62(5)$ & $80(30)$ & ps & $0.120$ & $1.80(68)$ & $0.120(23)$\\
\cellcolor{yellow!50} & \cellcolor{orange!50}$^{227}$Fr$_{2.47(3)\text{m}}$ & \cellcolor{gray!50} $3/2^+ \rightarrow \bm{1/2^+}$ & $39.9(4)$ & $2.7(2)$ & ns & $0.181$ & $2.39(34)$ & \cellcolor{blue!35} $0.14(1)$\\
\cellcolor{yellow!50} & \cellcolor{orange!50}$^{227}$Fr$^{(3,1)}_{2.47(3)\text{m}}$ & \cellcolor{gray!50} $3/2^+ \rightarrow \bm{1/2^+}$ & $144.16(4)$ & $38(12)$ & ps & $0.181$ & $2.30(76)$ & \cellcolor{blue!35} $0.13(2)$\\
\hline
\cellcolor{yellow!50}  $Z=88$ & \cellcolor{red!50} $^{227}$Ra$^{\dagger}_{42.2(5)\text{m}}$ & $1/2^+ \rightarrow \bm{3/2^+}$ & $120.711(4)$ & $\le47$ & ps & $0.181$ & $\ge5.1$ & $\ge0.202$\\ %$5.1(31)$ & $0.202(62)$\\
\hline
$Z=89$ & \cellcolor{magenta!50} $^{223}$Ac$^{\dagger}_{2.10(5)\text{m}}$ & $(7/2^-) \rightarrow \bm{(5/2^-)}$ & $42.4(1)$ & $\le250$ & ps & $0.147$ & $\ge4.5$ & $\ge0.187$\\ %$4.5(24)$ & $0.187(51)$\\
\hline
$Z=90$ & \cellcolor{magenta!50} $^{229}$Th$_{7880(120)\text{y}}$ & $7/2^+ \rightarrow \bm{5/2^+}$ & $42.4349(2)$ & $172(6)$ & ps & $0.190$ & $6.95(26)$ & \cellcolor{blue!35} $0.2293(44)$\\
\hline
$Z=92$ & $^{233}$U$^{(2,1)}_{1.5919(15)\times10^{5}\text{y}}$ & $7/2^+ \rightarrow \bm{5/2^+}$ & $40.351(7)$ & $110(80)$ & ps & $0.207$ & $10.0(74)$ & \cellcolor{blue!35} $0.27(1)$\\
& \cellcolor{red!50} $^{235}$U$^{\dagger(2)}_{7.04(1)\times10^{8}\text{y}}$ & $9/2^- \rightarrow \bm{7/2^-}$ & $46.103(8)$ & $\simeq14$ & ps & $0.215$ & $\simeq10$ & \cellcolor{blue!35} $\simeq0.27$\\%$10.2(51)$ & $0.270(67)$\\
\hline
$Z=93$ & \cellcolor{red!50} $^{237}$Np$^{(2,1)}_{2.144(7)\times10^{6}\text{y}}$ & $7/2^+ \rightarrow \bm{5/2^+}$ & $33.19629(22)$ & $54(24)$ & ps & $0.215$ & $10.8(49)$ & $0.274(62)$\\
\hline
%$Z=94$ & $^{239}$Pu$^{(N,1)}_{24110(30)\text{y}}$ & $3/2^+ \rightarrow \bm{1/2^+}$ & $7.861(2)$ & $36(3)$ & ps & $0.223$ & $12.6(10)$ & $0.291(11)$\\
$Z=95$ & \cellcolor{red!50} $^{243}$Am$^{\dagger}_{7364(22)\text{y}}$ & $7/2^- \rightarrow \bm{5/2^-}$ & $42.20(22)$ & $\simeq 40$ & ps & $0.224$ & $\simeq 13.6$ & $\simeq0.298$\\ %$13.6(69)$ & $0.298(75)$\\
\hline
$Z=97$ & \cellcolor{red!50} $^{249}$Bk$_{327.2(3)\text{d}}$ & $9/2^+ \rightarrow \bm{7/2^+}$ & $41.805(8)$ & $9(2)$ & ps & $0.235$ & $15.0(32)$ & \cellcolor{blue!35} $0.300(33)$\\
\hline
\hline
%\hhline{=========}
\end{longtable}
\end{footnotesize}

\section{Conclusion}

Interpreting atomic EDM measurements in terms of the EDM of the constituent particles \cite{Flambaum1994-fp,Lackenby2018-ua} or the mutual CP violating interactions \cite{Engel2013-rk,Chupp2015-ns} between them requires knowledge of nuclear deformation. While the theoretical models that predict the nuclear deformations are mature \cite{Agbemava2016-rx,Ebata2017-nt,Moller2016-gt}, there are however mutual inconsistencies between the various theoretical models, as well as between the models and measurements. This demands the explicit measurement of the nuclear deformations relevant for EDM measurements \cite{Mohanmurthy2020-np,MohanMurthy2024-dh}. Fortunately, E$2$ nuclear transitions to the ground state of various isotopes has already been measured \cite{nndc}. In this work we have used the existing data on E$2$ transitions to infer the quadrupole deformation, before setting out on a campaign to measure those where such measurements are not available a-priori.

There is an overlap in $32$ candidate isotopes between pure E$2$ transitions and mixed M$1$+E$2$ transitions. When considering a single type of transition, there could still be two or more different transitions within the same isotope, typically offset by $\Delta j = \pm 2$ in the case of E$2$ transitions and $\Delta j = \pm\{1,2,3\}$ in the case of M$1+$E$2$ transitions. The values for the quadrupole deformation we obtained through the study of these two different kinds of transitions, are remarkably consistent, demonstrating the versatility of the analysis. However, it is important to note that the uncertainty in our estimates of the quadrupole deformation is mainly dictated by the uncertainty associated with the lifetime of the transition, and to a negligible degree the uncertainty associated with the transition energy difference. The final uncertainties in the quadrupole deformation also have significant contribution from our Weisskopf estimates of the transition lifetime. Due to this, the uncertainty of the deformation estimates for mildly to moderately deformed species like $^{189,191,193}$Ir, $^{195}$Pt, and $^{223,225}$Ra are heavily subject to the Weisskopf estimates, and therefore underestimated. The final estimates have no contributions from the N-body interactions, and therefore are underestimated. We have also chosen to neglect the effects of nuclear surface diffuseness \cite{Swiatecki1955-rj}, that further modify the transition matrix elements in Eq.~\ref{eq2024-1-5}.

Furthermore, especially in the octupole deformation region, the quadrupole deformation of the isotopes of $^{229}$Th, $^{231}$Pa, and $^{239}$Pu, could not be precisely characterized due to interactions of the higher order deformations. In these species, the uncertainty associated with the quadrupole deformation has been under-estimated. In the isotope %$^{151}$Eu, $^{185}$Au, and
$^{195}$Pt, whose lifetimes are relatively well measured, the other effects we have neglected here contribute significantly, therefore the uncertainty in the extracted quadrupole deformation is underestimated.

The quadrupole deformation of certain isotopes could not be characterized in this method even though the data regarding their transition lifetimes was available. Isotopes of %$^{159}$Tb,
$^{151}$Eu, $^{167}$Er, $^{183}$W, $^{187,197}$Pt, $^{185,196,197,198}$Au, $^{195,197,199,201}$Hg, $^{200}$Tl, $^{219}$Rn, and $^{229}$Ra were left out since they all involved non-band energy levels in the transition measurements. In isotopes such as $^{152}$Eu, $^{160}$Tb, $^{171}$Hf, $^{177}$W, $^{186}$Re, $^{177}$Os and $^{185}$Ir, the transition lifetimes were unusually large, larger than that allowed by the Weisskopf estimates, indicating other effects in play, \emph{e.g.}~unavailability of low-lying spin states from which to E$2$ transition down to the ground state. %In isotopes of $^{151}$Tb and $^{187}$Pt, $^{171}$Er, $^{175}$Ta, $^{205,207}$Po, $^{239}$U, $^{241}$Pu, and $^{243}$Cm, other reasons maybe that certain low lying E$2$ and M$1$ transitions are forbidden \cite{Fivet2016-ye}.

Typically, the isotopes relevant for EDM searches \cite{Mohanmurthy2021-ou} have a spin non-zero ground state. Using the E$2$ or M$1+$E$2$ transition lifetime to their ground state, and comparing them against the E$2$ transition strength from the very well vetted measurements in the nearest even-even nuclei \cite{Pritychenko2016-tw,Pritychenko2017-pm}, $B(E2:2^+\rightarrow0^+)$, we were able to infer the quadrupole deformation of $67$ isotopes in the mass range of $150\le A \le250$. Such a comparison was made possible since the spin of most of our candidate nuclei, the vast majority of which had a ground state spins greater than $0$, was carried by a single valence nucleon, leaving a core of $0$ spin \cite{Morse2020-ex}. Yet, we showed that in such cases special care must be taken with the application of appropriate correction factors (Tables A.1-A.3 in \emph{Ref.}~\cite{van-Dommelen2012-fg}), including the spin-multiplicity according to Eq.~\ref{eq2024-1-5}. We achieved this comprehensive characterization of quadrupole deformation by carefully considering the single particle contributions to the strength of the E$2$ transitions, as well as M$1$ transitions, which in turn allowed us to isolate the contribution due to quadrupole deformation. The studies of single particle Weisskopf estimates were previously lacking in this mass range. Our work now paves the way and provides the much needed impetus to precisely measure the nuclear deformation in species like $^{223,225}$Fr, $^{225,227}$Ac and $^{229}$Pa, where EDM measurements are foreseen \cite{Mohanmurthy2020-np,MohanMurthy2024-dh}, but no measurement of nuclear deformations exists.

\section*{Acknowledgement}

We would like to acknowledge useful discussions with Prof.~Dr.~Anatoli Afanasjev and Prof.~Dr.~Robert Redwine. P.M. and J.A.W were supported by US-DOE grant \#DE-SC0014448, P.M. was supported in addition by US-DOE grants \#DE-SC0018229 and DE-SC0019768. L.Q. was supported by MIT LNS and by Prof.~Dr.~Robert Redwine.


\begin{thebibliography}{99}

\bibitem{Riotto1999-vo}{A. Riotto and M. Trodden, {\em Recent Progress In Baryogenesis}, Annu. Rev. Nucl. Part. Sci. {\bf 49}, 35 (1999). DOI: \href{http://dx.doi.org/10.1146/annurev.nucl.49.1.35}{10.1146/annurev.nucl.49.1.35}.}

\bibitem{Morrissey2012-tt}{D. E. Morrissey and M. J. Ramsey-Musolf, {\em Electroweak Baryogenesis}, New J. Phys. {\bf 14}, 125003 (2012). DOI: \href{http://dx.doi.org/10.1088/1367-2630/14/12/125003}{10.1088/1367-2630/14/12/125003}.}

\bibitem{Aghanim2020-cm}{N. Aghanim et al., {\em Planck 2018 Results - VI. Cosmological Parameters}, Astron. Astrophys. Suppl. Ser. {\bf 641}, A6 (2020). DOI: \href{http://dx.doi.org/10.1051/0004-6361/201833910}{10.1051/0004-6361/201833910}.}

\bibitem{Chupp2015-ns}{T. Chupp and M. Ramsey-Musolf, {\em Electric Dipole Moments: A Global Analysis}, Phys. Rev. C {\bf 91}, 035502 (2015). DOI: \href{http://dx.doi.org/10.1103/PhysRevC.91.035502}{10.1103/PhysRevC.91.035502}.}

\bibitem{t_Hooft1976-lq}{G. ’t Hooft, {\em Symmetry Breaking through Bell-Jackiw Anomalies}, Phys. Rev. Lett. {\bf 37}, 8 (1976). DOI: \href{http://dx.doi.org/10.1103/PhysRevLett.37.8}{10.1103/PhysRevLett.37.8}.}

\bibitem{Kirch2020-dr}{K. Kirch and P. Schmidt-Wellenburg, {\em Search for Electric Dipole Moments}, Proc. of International Workshop on Flavour Changing and Conserving Processes (FCCP 2019), EPJ Web of Conf. {\bf 234}, 1007 (2020). DOI: \href{https://doi.org/10.1051/epjconf/202023401007}{10.1051/epjconf/202023401007}.}

\bibitem{Mohanmurthy2021-ou}{P. Mohanmurthy and J. A. Winger, {\em Estimation of CP Violating EDMs from Known Mechanisms in the SM}, Proc. of 40$^{th}$ International Conference on High Energy Physics (ICHEP2020), Proc. of Sci. {\bf 390}, 265 (2021).DOI: \href{http://dx.doi.org/10.22323/1.390.0265}{10.22323/1.390.0265}. \href{https://arxiv.org/abs/2009.00852}{arXiv: [2009.00852].}}

\bibitem{Engel2013-rk}{J. Engel, M. J. Ramsey-Musolf, and U. van Kolck, {\em Electric Dipole Moments of Nucleons, Nuclei, and Atoms: The Standard Model and Beyond}, Prog. Part. Nucl. Phys. {\bf 71}, 21 (2013). DOI: \href{http://dx.doi.org/10.1016/j.ppnp.2013.03.003}{10.1016/j.ppnp.2013.03.003}.}

\bibitem{Chupp2019-cm}{T. E. Chupp, P. Fierlinger, M. J. Ramsey-Musolf, and J. T. Singh, {\em Electric Dipole Moments of Atoms, Molecules, Nuclei, and Particles}, Rev. Mod. Phys. {\bf 91}, 015001 (2019). DOI: \href{http://dx.doi.org/10.1103/RevModPhys.91.015001}{10.1103/RevModPhys.91.015001}.}

\bibitem{Flambaum1994-fp}{V. V. Flambaum, {\em Spin Hedgehog and Collective Magnetic Quadrupole Moments Induced by Parity and Time Invariance Violating Interaction}, Phys. Lett. B {\bf 320}, 211 (1994). DOI: \href{http://dx.doi.org/10.1016/0370-2693(94)90646-7}{10.1016/0370-2693(94)90646-7}.}

\bibitem{Flambaum2014-qn}{V. V. Flambaum, D. DeMille, and M. G. Kozlov, {\em Time-Reversal Symmetry Violation in Molecules Induced by Nuclear Magnetic Quadrupole Moments}, Phys. Rev. Lett. {\bf 113}, 103003 (2014). DOI: \href{http://dx.doi.org/10.1103/PhysRevLett.113.103003}{10.1103/PhysRevLett.113.103003}.}

\bibitem{Sushkov1984-cb}{O. P. Sushkov, V. V. Flambaum, and I. B. Khriplovich, {\em Possibility of Investigating P-and T-Odd Nuclear Forces in Atomic and Molecular Experiments}, JETP Letters {\bf 60}, 873 (1984). URL: \href{https://inis.iaea.org/search/search.aspx?orig_q=RN:17005522}{inis.iaea.org/search/search.aspx?orig\_q=RN:17005522}.}

\bibitem{Auerbach1996-sp}{N. Auerbach, V. V. Flambaum, and V. Spevak, {\em Collective T- and P-Odd Electromagnetic Moments in Nuclei with Octupole Deformations}, Phys. Rev. Lett. {\bf 76}, 4316 (1996). DOI: \href{http://dx.doi.org/10.1103/PhysRevLett.76.4316}{10.1103/PhysRevLett.76.4316}.}

\bibitem{Spevak1997-bg}{V. Spevak, N. Auerbach, and V. V. Flambaum, {\em Enhanced T-Odd, P-Odd Electromagnetic Moments in Reflection Asymmetric Nuclei}, Phys. Rev. C {\bf 56}, 1357 (1997). DOI: \href{http://dx.doi.org/10.1103/PhysRevC.56.1357}{10.1103/PhysRevC.56.1357}.}

\bibitem{Dzuba2002-or}{V. A. Dzuba, V. V. Flambaum, J. S. M. Ginges, and M. G. Kozlov, {\em Electric Dipole Moments of Hg, Xe, Rn, Ra, Pu, and TlF Induced by the Nuclear Schiff Moment and Limits on Time-Reversal Violating Interactions}, Phys. Rev. A {\bf 66}, 012111 (2002). DOI: \href{http://dx.doi.org/10.1103/PhysRevA.66.012111}{10.1103/PhysRevA.66.012111}.}

\bibitem{Khriplovich2012-if}{I. B. Khriplovich and S. K. Lamoreaux, {\em CP Violation Without Strangeness: Electric Dipole Moments of Particles, Atoms, and Molecules}, Springer Science \& Business Media (2012). URL: \href{http://inspirehep.net/record/460248?ln=en}{inspirehep.net/record/460248}.}

\bibitem{Lackenby2018-ua}{B. G. C. Lackenby and V. V. Flambaum, {\em Time Reversal Violating Magnetic Quadrupole Moment in Heavy Deformed Nuclei}, Phys. Rev. D {\bf 98}, 115019 (2018). DOI: \href{http://dx.doi.org/10.1103/PhysRevD.98.115019}{10.1103/PhysRevD.98.115019}.}

\bibitem{Flambaum2022-kj}{V. V. Flambaum and A. J. Mansour, {\em Enhanced Magnetic Quadrupole Moments in Nuclei with Octupole Deformation and Their CP-Violating Effects in Molecules}, Phys. Rev. C {\bf 105}, 065503 (2022). DOI: \href{http://dx.doi.org/10.1103/PhysRevC.105.065503}{10.1103/PhysRevC.105.065503}.}

\bibitem{Mohanmurthy2020-np}{P. Mohanmurthy, U. Silwal, D. P. Siwakoti, and J. A. Winger, {\em Survey of Deformation in Nuclei in Order to Estimate the Enhancement of Sensitivity to Atomic EDM}, Proc. of 15$^{th}$ International Conference on Meson-Nucleon Physics and the Structure of the Nucleon (MENU 2019), AIP Conf. Proc. {\bf 2249}, 030046 (2020). DOI: \href{http://dx.doi.org/10.1063/5.0008560}{10.1063/5.0008560}. \href{https://arxiv.org/abs/1912.09271}{arXiv: [1912.09271]}}

\bibitem{MohanMurthy2024-dh}{P. MohanMurthy, U. Silwal, and J. A. Winger, {\em A Survey of Nuclear Quadrupole Deformation in Order to Estimate the Nuclear MQM and Its Relative Contribution to the Atomic EDM}, Proc. of International Conference on Hyperfine Interactions and their Applications (HYPERFINE 2023), Interactions {\bf 245}, 64 (2024). DOI: \href{http://dx.doi.org/10.1007/s10751-024-01880-7}{10.1007/s10751-024-01880-7}. \href{https://arxiv.org/abs/2401.11055}{arXiv: [2401.11055]}.}

\bibitem{Leander1986-fr}{G. A. Leander, W. Nazarewicz, G. F. Bertsch, and J. Dudek, {\em Low-Energy Collective E1 Mode in Nuclei}, Nucl. Phys. A {\bf 453}, 58 (1986). DOI: \href{http://dx.doi.org/10.1016/0375-9474(86)90029-1}{10.1016/0375-9474(86)90029-1}.}

\bibitem{Pritychenko2016-tw}{B. Pritychenko, M. Birch, B. Singh, and M. Horoi, {\em Tables of E2 Transition Probabilities from the First $2^+$ States in Even-Even Nuclei}, At. Data Nucl. Data Tables {\bf 107}, 1 (2016). DOI: \href{http://dx.doi.org/10.1016/j.adt.2015.10.001}{10.1016/j.adt.2015.10.001}.}

\bibitem{Pritychenko2017-pm}{B. Pritychenko, M. Birch, and B. Singh, {\em Revisiting Grodzins Systematics of B(E2) Values}, Nucl. Phys. A {\bf 962}, 73 (2017). DOI: \href{http://dx.doi.org/10.1016/j.nuclphysa.2017.03.011}{10.1016/j.nuclphysa.2017.03.011}.}

\bibitem{nndc}{National Nuclear Data Center (NNDC). URL: \href{https://www.nndc.bnl.gov/}{www.nndc.bnl.gov}. Accessed: Aug 1, 2024.}

\bibitem{Garg2023-jx}{S. Garg, B. Maheshwari, B. Singh, Y. Sun, A. Goel, and A. K. Jain, {\em Atlas of Nuclear Isomers - Second Edition}, At. Data Nucl. Data Tables {\bf 150}, 101546 (2023). DOI: \href{http://dx.doi.org/10.1016/j.adt.2022.101546}{10.1016/j.adt.2022.101546}.}

\bibitem{Heyde2011-mc}{K. Heyde and J. L. Wood, {\em Shape Coexistence in Atomic Nuclei}, Rev. Mod. Phys. {\bf 83}, 1467 (2011). DOI: \href{http://dx.doi.org/10.1103/RevModPhys.83.1467}{10.1103/RevModPhys.83.1467}.}

\bibitem{Dracoulis2016-uf}{G. D. Dracoulis, P. M. Walker, and F. G. Kondev, {\em Review of Metastable States in Heavy Nuclei}, Rep. Prog. Phys. {\bf 79}, 076301 (2016). DOI: \href{http://dx.doi.org/10.1088/0034-4885/79/7/076301}{10.1088/0034-4885/79/7/076301}.}

\bibitem{Agbemava2016-rx}{S. E. Agbemava, A. V. Afanasjev, and P. Ring, {\em Octupole Deformation in the Ground States of Even-Even Nuclei: A Global Analysis within the Covariant Density Functional Theory}, Phys. Rev. C {\bf 93}, 044304 (2016). DOI: \href{http://dx.doi.org/10.1103/PhysRevC.93.044304}{10.1103/PhysRevC.93.044304}.}

\bibitem{Ebata2017-nt}{S. Ebata and T. Nakatsukasa, {\em Octupole Deformation in the Nuclear Chart Based on the 3D Skyrme Hartree-Fock plus BCS Model}, Phys. Scr. {\bf 92}, 064005 (2017). DOI: \href{http://dx.doi.org/10.1088/1402-4896/aa6c4c}{10.1088/1402-4896/aa6c4c}.}

\bibitem{Moller2016-gt}{P. M\"oller, A. J. Sierk, T. Ichikawa, and H. Sagawa, {\em Nuclear Ground-State Masses and Deformations: FRDM(2012)}, At. Data Nucl. Data Tables {\bf 1}, 109-110 (2016). DOI: \href{http://dx.doi.org/10.1016/j.adt.2015.10.002}{10.1016/j.adt.2015.10.002}.}

\bibitem{Blatt1979-vo}{J. M. Blatt and V. F. Weisskopf, {\em Interaction of Nuclei with Electromagnetic Radiation, Theoretical Nuclear Physics}, Springer New York, 583-669 (1979). DOI: \href{http://dx.doi.org/10.1007/978-1-4612-9959-2_12}{10.1007/978-1-4612-9959-2\_12}.}

\bibitem{Blatt2012-fk}{J. M. Blatt and V. F. Weisskopf, {\em Theoretical Nuclear Physics}, 1979th Ed., Springer New York (2012). DOI: \href{http://dx.doi.org/10.1007/978-1-4612-9959-2}{10.1007/978-1-4612-9959-2}.}

\bibitem{Weisskopf1957-sa}{V. F. Weisskopf, {\em Nuclear Physics}, Rev. Mod. Phys. {\bf 29}, 174 (1957). DOI: \href{http://dx.doi.org/10.1103/RevModPhys.29.174}{10.1103/RevModPhys.29.174}.}

\bibitem{Moszkowski1953-ok}{S. A. Moszkowski, {\em Lifetimes of Nuclear Isomers}, Phys. Rev. {\bf 89}, 474 (1953). DOI: \href{http://dx.doi.org/10.1103/PhysRev.89.474}{10.1103/PhysRev.89.474}.}

\bibitem{Krane1987-lw}{K. S. Krane, {\em Introductory Nuclear Physics}, 3rd ed., John Wiley \& Sons (1987). URL: \href{https://www.wiley.com/en-us/Introductory+Nuclear+Physics\%2C+3rd+Edition-p-9780471805533}{www.wiley.com/en-us/Introductory+Nuclear+Physics\%2C+3rd+Edition-p-9780471805533}.}

\bibitem{Miyagi2024-iy}{T. Miyagi, X. Cao, R. Seutin, S. Bacca, R. F. G. Ruiz, K. Hebeler, J. D. Holt, and A. Schwenk, {\em Impact of Two-Body Currents on Magnetic Dipole Moments of Nuclei}, Phys. Rev. Lett. {\bf 132}, 232503 (2024). DOI: \href{http://dx.doi.org/10.1103/PhysRevLett.132.232503}{10.1103/PhysRevLett.132.232503}.}

\bibitem{Stroberg2022-uk}{S. R. Stroberg, J. Henderson, G. Hackman, P. Ruotsalainen, G. Hagen, and J. D. Holt, {\em Systematics of E2 Strength in the $sd$ Shell with the Valence-Space in-Medium Similarity Renormalization Group}, Phys. Rev. C. {\bf 105}, 034333 (2022). DOI: \href{http://dx.doi.org/10.1103/PhysRevC.105.034333}{10.1103/PhysRevC.105.034333}.}

\bibitem{Wollersheim1993-ro}{H. J. Wollersheim et al., {\em Coulomb Excitation of $^{226}$Ra}, Nucl. Phys. A {\bf 556}, 261 (1993). DOI: \href{http://dx.doi.org/10.1016/0375-9474(93)90351-W}{10.1016/0375-9474(93)90351-W}.}

\bibitem{Butler1996-px}{P. A. Butler and W. Nazarewicz, {\em Intrinsic Reflection Asymmetry in Atomic Nuclei}, Rev. Mod. Phys. {\bf 68}, 349 (1996). DOI: \href{http://dx.doi.org/10.1103/RevModPhys.68.349}{10.1103/RevModPhys.68.349}.}

\bibitem{Leander1988-ww}{G. A. Leander and Y. S. Chen, {\em Reflection-Asymmetric Rotor Model of Odd A$\sim$219-229 Nuclei}, Phys. Rev. C {\bf 37}, 2744 (1988). DOI: \href{http://dx.doi.org/10.1103/PhysRevC.37.2744}{10.1103/PhysRevC.37.2744}.}

\bibitem{Endt1979-sz}{P. M. Endt, {\em Strengths of Gamma-Ray Transitions in A = 45-90 Nuclei}, At. Data Nucl. Data Tables {\em 23}, 547 (1979). DOI: \href{http://dx.doi.org/10.1016/0092-640X(79)90030-5}{10.1016/0092-640X(79)90030-5}.}

\bibitem{Endt1993-fc}{P. M. Endt, {\em Strengths of Gamma-Ray Transitions in A = 5-44 Nuclei, IV}, At. Data Nucl. Data Tables {\bf 55}, 171 (1993). DOI: \href{http://dx.doi.org/10.1006/adnd.1993.1020}{10.1006/adnd.1993.1020}.}

\bibitem{Endt1981-wp}{P. M. Endt, {\em Strengths of Gamma-Ray Transitions in A = 91-150 Nuclei}, At. Data Nucl. Data Tables {\bf 26}, 47 (1981). DOI: \href{http://dx.doi.org/10.1016/0092-640X(81)90011-5}{10.1016/0092-640X(81)90011-5}.}

\bibitem{Siegbahn1965-le}{K. Siegbahn, {\em Alpha-, Beta- and Gamma-Ray Spectroscopy}, Vol. 2, North-Holland Amsterdam (1965). URL: \href{https://search.worldcat.org/title/Alpha-beta-and-gamma-ray-spectroscopy.-vol.-2/oclc/768663712}{search.worldcat.org/title/Alpha-beta-and-gamma-ray-spectroscopy.-vol.-2/oclc/768663712}.}

\bibitem{Stech1952-np}{B. Stech, {\em Die Lebensdauer Isomerer Kerne}, Z. Naturforsch. A {\bf 7}, 401 (1952). DOI: \href{http://dx.doi.org/10.1515/zna-1952-0604}{10.1515/zna-1952-0604}.}

\bibitem{van-Dommelen2012-fg}{L. van Dommelen, {\em Quantum Mechanics for Engineers}, Florida State University, Tallahassee, FL (2012). URL: \href{https://www.semanticscholar.org/paper/Quantum-mechanics-for-engineers.-Dommelen/2eb883af5ccb0d0bdd9f0d10009c4a7151c9c339}{semanticscholar.org/paper/Quantum-mechanics-for-engineers.-Dommelen/2eb883af5ccb0d0bdd9f0d10009c4a7151c9c339}.}

\bibitem{McClelland2016-od}{J. J. McClelland, A. V. Steele, B. Knuffman, K. A. Twedt, A. Schwarzkopf, and T. M. Wilson, {\em Bright Focused Ion Beam Sources Based on Laser-Cooled Atoms}, Appl Phys Rev {\bf 3}, (2016). DOI: \href{http://dx.doi.org/10.1063/1.4944491}{10.1063/1.4944491}.}

\bibitem{Swiatecki1955-rj}{W. J. Swiatecki, {\em Nuclear Surface Energy and the Diffuseness of the Nuclear Surface}, Phys. Rev. {\bf 98}, 203 (1955). DOI: \href{http://dx.doi.org/10.1103/PhysRev.98.203}{10.1103/PhysRev.98.203}.}

%\bibitem{Fivet2016-ye}{V. Fivet, P. Quinet, and M. A. Bautista, {\em Radiative Rates for Forbidden M1 and E2 Transitions of Astrophysical Interest in Doubly Ionized Iron-Peak Elements}, Astron. Astrophys. {\bf 585}, A121 (2016). DOI: \href{http://dx.doi.org/10.1051/0004-6361/201526983}{10.1051/0004-6361/201526983}.}

\bibitem{Morse2020-ex}{C. Morse et al., {\em Quadrupole and Octupole Collectivity in $^{143}$Ba}, Phys. Rev. C {\bf 102}, 054328 (2020). DOI: \href{http://dx.doi.org/10.1103/PhysRevC.102.054328}{10.1103/PhysRevC.102.054328}.}



\end{thebibliography}
\end{document}